\documentclass[lettersize,journal]{IEEEtran} \def\TwoColumn{1}

\usepackage{xcolor}
\usepackage{amsmath,amssymb,amsfonts,amsthm}
\usepackage{algorithmic}
\usepackage{graphicx}
\usepackage{dsfont}
\usepackage{pstricks}
\usepackage{afterpage}
\usepackage{comment}
\usepackage{cite}
\usepackage{url}
\usepackage{etoolbox}
\usepackage{hyperref}
\hypersetup{hidelinks=true}
\usepackage{textcomp}
\def\BibTeX{{\rm B\kern-.05em{\sc i\kern-.025em b}\kern-.08em
    T\kern-.1667em\lower.7ex\hbox{E}\kern-.125emX}}
\allowdisplaybreaks[1]
\usepackage{tikz}
\usetikzlibrary{arrows}
\usetikzlibrary{shapes}

\newcommand*\circled[1]{\tikz[baseline=(char.base)]{
            \node[shape=circle,draw,inner sep=0.7pt] (char) {#1};}}

\ifdef{\TwoColumn}{}{ \usepackage{setspace}\doublespacing}

\makeatletter
\def\@endtheorem{$\blacksquare$\endtrivlist\@endpefalse } %
\makeatother

\newtheoremstyle{thmstyle}%
{}%
{}%
{}%
{}%
{\bfseries}%
{.}%
{ }%
{}%
\theoremstyle{thmstyle}

\usepackage[overload]{textcase}

\usepackage{textcomp}
\def\BibTeX{{\rm B\kern-.05em{\sc i\kern-.025em b}\kern-.08em
    T\kern-.1667em\lower.7ex\hbox{E}\kern-.125emX}}

\begin{document}
\title{Real-Time Multi-Modal Subcomponent-Level Measurements for Trustworthy System
  Monitoring and Malware Detection}
\author{F. Khorrami, R. Karri, P. Krishnamurthy
	\thanks{F. Khorrami, R. Karri, and P. Krishnamurthy are with the
		Dept. of ECE, NYU Tandon School of Engineering, Brooklyn, NY 11201, USA.
		(e-mails: \{khorrami, rkarri, prashanth.krishnamurthy\}@nyu.edu).}
	\thanks{This work is supported in part by the Defense Advanced Research Projects Agency (DARPA) under contract HR00112390029. The views and conclusions contained in this document are those of the authors and should not be interpreted as representing the official policies, either expressed or implied, of the U.S. Government.}
}

\maketitle

\begin{abstract}
	With increasingly sophisticated cyber-adversaries able to access a wider
  repertoire of mechanisms to implant malware such as ransomware, CPU/GPU
  keyloggers, and stealthy kernel rootkits, there is an urgent need for
  techniques to detect and mitigate such attacks. While state of the art in
  anomaly detection in modern computers relies on digital and analog side
  channel measurements assuming trustworthiness of measurements obtained on the
  main processor, such an approach has limitations since processor-based side
  channel measurements are potentially untrustworthy. For example, sophisticated
  adversaries (especially in late stage cyber attacks when they have breached
  the computer and network security systems such as firewalls and antivirus  and
  penetrated the computer's operating system) can compromise user-space and
  kernel-space measurements. To address this key limitation of state of the art,
  we propose instead a ``subcomponent-level'' approach to collect side channel
  measurements  so as to enable robust anomaly detection in a modern computer
  even when the main processor is compromised. In particular, our proposed
  approach leverages the fact that modern computers are complex systems with
  multiple interacting subcomponents and measurements from these subcomponents
  can be used to detect anomalies even when the main processor is no longer
  trustworthy under an attack. For this purpose, we develop mechanisms to obtain
  time series measurements of activity of several subcomponents and
  methodologies to process and fuse these measurements for the purpose of
  anomaly detection. The subcomponents addressed include network interface
  controller (NIC), Graphics Processing Unit (GPU), CPU Hardware Performance Counters, CPU power, and keyboard. Our main hypothesis
  is that subcomponent measurements can enable detection of security threats
  without requiring a trustworthy main processor. By enabling real-time
  measurements from multiple subcomponents, the goal of the project is to
  provide a deeper visibility into the operation of the system, thereby yielding
  a powerful tool to track system operation and detect anomalies. 
\end{abstract}


\section{Introduction}
State of the art in anomaly detection in computing systems relies on digital/analog side channels (e.g., Hardware Performance Counters, network communication and power consumption). Limitations of such methods are twofold. Firstly, measurements obtained using software on the processor are potentially untrustworthy since an adversary with access to the system can have the ability to corrupt readings obtained from user space or kernel measurements. Secondly, side channel measurements have limits as to their level of detail/granularity in terms of identifying behavior of individual subcomponents of the system and thereby in enabling detection of anomalies that manifest themselves in terms of the activity patterns of different subcomponents of the system.
The need for robust anomaly detection methodologies is becoming increasingly crucial \cite{kkk16,nadim2023kernellevel} due to
increasingly sophisticated cyber-adversaries able to access a wider repertoire of mechanisms to implant malware such as ransomware, CPU/GPU keyloggers, and stealthy kernel rootkits~\cite{krishnamurthy2019stealthy}. Hence, there is an urgent need for techniques to detect and mitigate such attacks.
Sophisticated adversaries (especially in late stage cyber attacks when they have breached the computer and network security systems such as firewalls and antivirus and penetrated the computer's operating system) can compromise user-space and kernel-space measurements.
Hence, methods that base their anomaly detection on digital and analog side channel measurements acquired assuming trustworthiness of measurements obtained on the main processor can be rendered ineffective since a late-stage adversary can manipulate processor-based side channel measurements and make them potentially untrustworthy.

To address the key limitations outlined above of the current state of the art, the proposed approach addresses the development of methodologies to enable robust anomaly detection in a modern computer even when the main processor is compromised.
In particular, the proposed approach leverages the fact that modern computers are complex systems with multiple interacting subcomponents communicating with each other over various buses and measurements obtained directly from these subcomponents/buses can be used to detect anomalies even when the main processor is no longer trustworthy under an attack.
For this purpose, we developed mechanisms to obtain time series measurements of subcomponent activity and methodologies to process and fuse these measurements for the purpose of anomaly detection.
We consider multiple subcomponents that are typically part of a modern computer including the network interface controller (NIC), Graphics Processing Unit (GPU),  keyboard, CPU Hardware Performance Counters, CPU power measurements, and SATA controller.
Our main hypothesis is that subcomponent measurements can enable detection of security threats without requiring a trustworthy main processor.

\begin{figure*}[!t]
\centering
\includegraphics[width=0.8\textwidth]{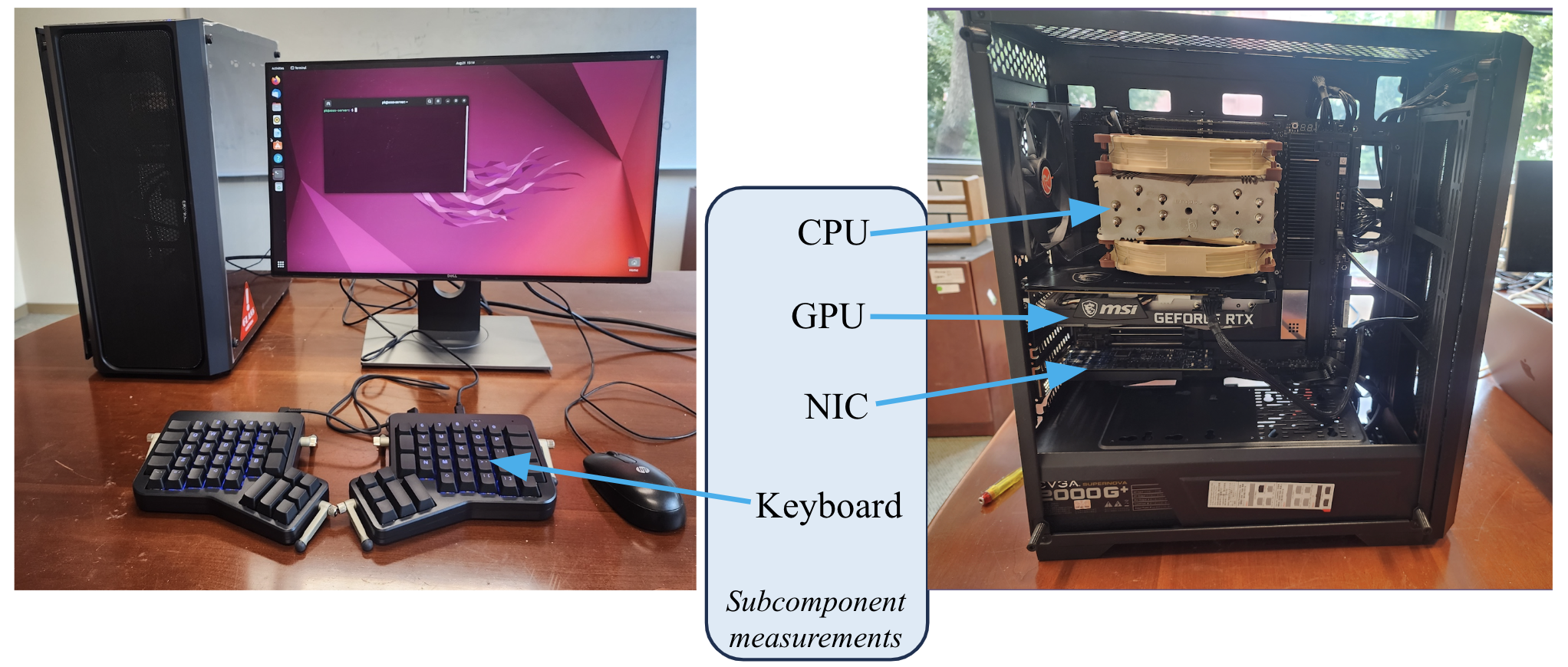}
\vspace*{-0.01in}
\caption{Integrated testbed for collection of subcomponent measurement datasets.}
\label{fig:integrated_testbed}
\vspace*{-0.01in}
\end{figure*}

By enabling real-time measurements from multiple subcomponents, the goal of the proposed approach is to provide a deeper visibility into the operation of the overall computer system, thereby yielding a powerful tool to track system operation and detect anomalies, especially focusing on scenarios involving late-stage cyber-attacks. A major focus of the effort was to collect time-synchronized time series data from multiple subcomponents while running various types of malware.
For this purpose, an integrated experimental testbed (Section~\ref{sec:testbed}) was developed to enable
simultaneous data collection from multiple subcomponents integrated into the
experimental testbed (Figure~\ref{fig:integrated_testbed}) as described in
Section~\ref{sec:subcomponents}.
Since the SATA data collection currently runs as part of a separate execution
framework, SATA-based measurements were not collected as part of this dataset.
However, the separate framework based on SATA and NBD (Network Block Device)
enables deep filesystem-aware side channels that can provide complementary
benefits along with the other side channels in this dataset. Hence, this
separate framework is discussed in Appendix~\ref{sec:sata_nbd}.
Sample analyses of the subcomponent
measurements for anomaly detection is discussed in Section~\ref{sec:analysis}.

\begin{figure*}[!t]
\centerline{
\includegraphics[width=0.9\textwidth]{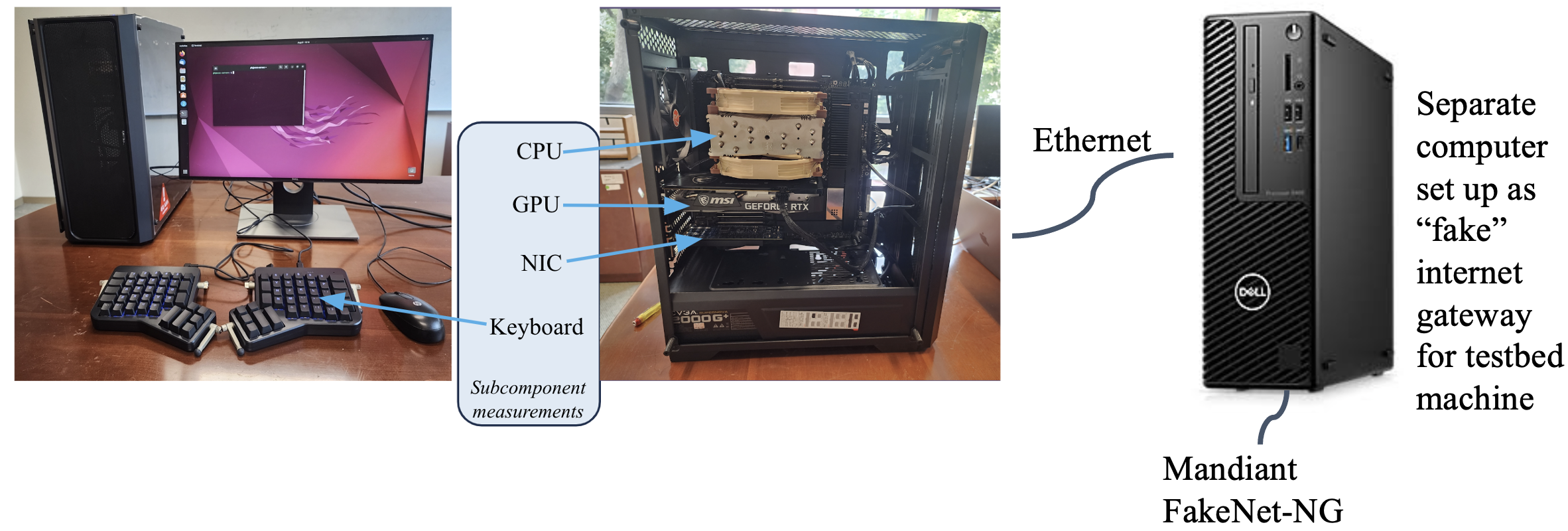}
}
\vspace*{-0.01in}
\caption{Integrated testbed for subcomponent measurements with Mandiant FakeNet-NG used to emulate connectivity from the testbed machine.}
\label{fig:integrated_testbed_overall_fakenet}
\vspace*{-0.01in}
\end{figure*}

\section{Experimental Testbed and Dataset Collection Framework}
\label{sec:testbed}
The integrated testbed is shown in Figure~\ref{fig:integrated_testbed}. The testbed computer uses an
Intel Xeon W5-3435X processor with an
Asus W790E motherboard, which supports DDR5 memory and up to 7 PCIe5.0x16
devices. Subcomponent measurement mechanisms were integrated for NIC (using a
Netronome Agilio CX 40 Gigabit Ethernet SmartNIC), GPU, keyboard, and CPU
(including HPCs and power measurements). Additionally, a filesystem-aware
storage activity measurement
framework was also integrated based on SATA and NBD as discussed further
in Appendix~\ref{sec:sata_nbd}. 

A primary focus in the development of the testbed and associated malware
execution and data collection framework was to enable a high level of automation
in running diverse malware samples and collecting measurement time series
datasets before, during, and after running the malware samples. One aspect of the integrated testbed and automated framework for data collection that we considered a few options for was the network connectivity. Several malware samples attempt to connect to an external server such as their C2
    (command and control) server, often as the one of the first actions on launching a malware sample. Many malware samples stop running if an available network connection is not detected. On the other hand, attempting to connect the testbed to the internet directly poses multiple issues such as the risk of spreading malware to other computers and the fact that our university network security systems would typically detect suspicious network activity (e.g., communications to known C2 servers) and block the network connection. One option we initially considered  was to use a mobile phone as the internet gateway.
Specifically, to facilitate running malware
    samples without being blocked by the university's network security systems while also
    being able to record network activity using the NIC on which we are running
    the custom-modified firmware, the test machine was connected via the NIC to a
    separate computer, which then was connected over WiFi to a mobile phone that functions as a mobile 5G hotspot.
This scheme enabled the malware to be able to reach the internet and the malware's C2 servers without being flagged by the university's network security systems which otherwise would block the internet access of the computer.
Since the first action of some malware samples is to try to
    communicate with their C2 servers, ensuring that these connections are not
    blocked is required to enable the malware to properly launch.
This allowed running of various types of malware including ones that were
flagged by the university's network security systems.  
 However, this scheme of using a mobile phone for internet connectivity had the limitation that the C2 servers for some malware samples were no longer accessible (e.g., no longer online at their original addresses, blocked by the mobile phone provider, etc.). Also, it would be useful to be able to control or dynamically influence the network communication with the C2 server. To address these issues with using a mobile phone for internet connectivity, the second option that we explored uses the Mandiant FakeNet-NG to emulate internet connectivity from the testbed machine (Figure~\ref{fig:integrated_testbed_overall_fakenet}). FakeNet-NG is an open-source software that can simulate a variety of network services including HTTP, HTTPS, DNS, FTP, and SMTP. This approach enabled running of malware samples without the need for a mobile phone for internet connectivity and without the risk of spreading malware to other computers. The FakeNet-NG software was run on a separate computer connected to the testbed machine via the NIC on which our custom firmware-level measurements are implemented. FakeNet-NG was configured to provide valid responses to several protocols, making it appear to the malware running on the testbed as if it is successfully accessing the internet.

To enable running of a wide range of malware samples that have unknown side
effects, a virtual machine (VM) approach was utilized. Using the VirtualBox API
functions to clone from a known-good VM, launch the VM and the malware sample
within it at precise pre-specified times, attempt to stop the malware, and
terminate the data collection and the VM, an automated framework for testing of
malware samples and collection of measurement datasets was developed.
The VM-based approach is more suitable (compared to running malware samples
natively or via a technique such as Docker) for diverse malware with unknown
side effects since it provides a higher level of isolation and control over the
malware execution environment with the VM image restored between runs.
The automated framework facilitates large-scale data collection from the integrated testbed while running malware samples (with precisely orchestrated VM environment setup and data collection start, malware launch, and malware kill stages during each sample run) and collecting time series measurements from the subcomponents. For each malware sample, the time series measurement datasets include time intervals before, during, and after running the malware sample. 

\begin{figure*}[!t]
  \centerline{
    \includegraphics[width=0.8\textwidth]{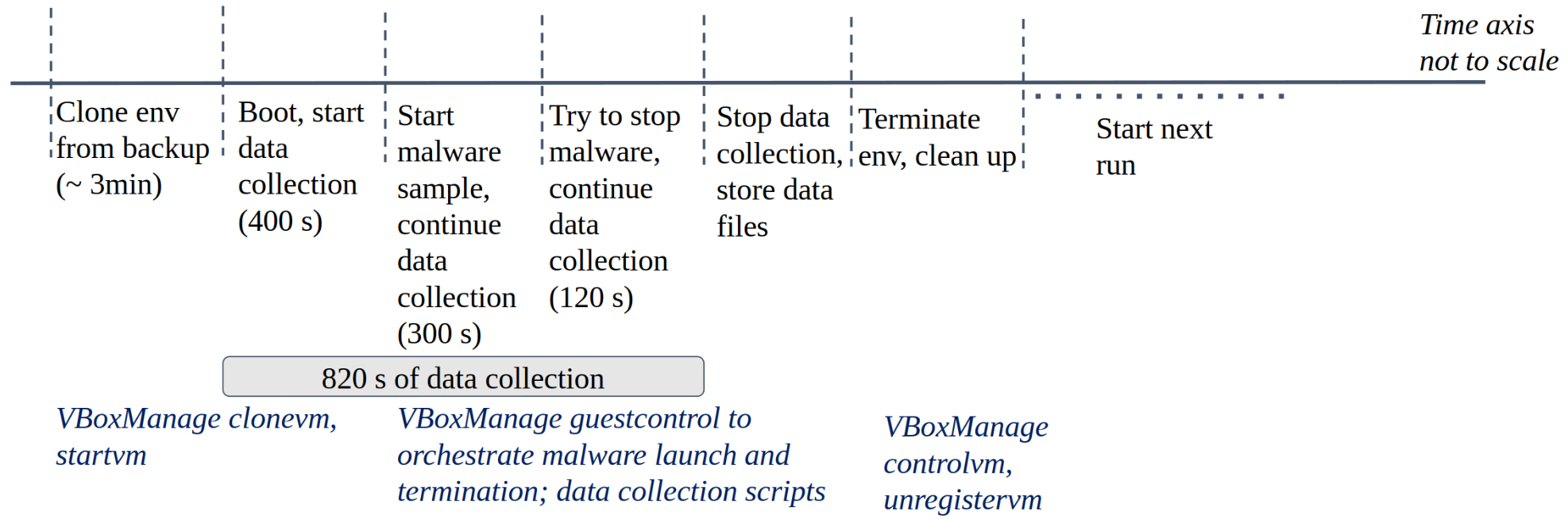}}
    \caption{Settings for time intervals for malware run and data collection on integrated testbed.}
  \label{fig:malware_testing_timeline}
\end{figure*}

To ensure a uniform and synchronized format for the datasets collected for the
different malware samples, the time settings for the malware run and data
collection are kept the same across all data collection runs. The time settings
are summarized in Figure~\ref{fig:malware_testing_timeline}. The dataset
includes measurements before running the malware sample, during the run of the
malware sample, and after attempting to terminate the malware sample using a
kill command (since some malware samples spawn other processes and/or infiltrate
the kernel via a rootkit, the malware behavior often continues after the kill
command). For each malware sample, data is collected for a total of 820 s as
shown in Figure~\ref{fig:malware_testing_timeline}. The VM based approach
enables running of malware samples with unknown side effects and ensures that
the environment is clean before starting the collection of data for a new
malware sample.

Using the experimental testbed and associated malware
execution and data collection framework discussed above, a measurement dataset
was collected for several different types of malware samples.
For each malware sample, the subcomponent measurements are collected as time series into csv files with a synchronized time base.  The timestamps are synchronized across all the subcomponents to facilitate multimodal analysis and anomaly detection.

\section{Subcomponent-Level Measurements}
\label{sec:subcomponents}
The subcomponent measurement mechanisms for each of the considered subcomponents are discussed in the subsections below.

\subsection{NIC (Agilio)}
\begin{figure*}[!t]
    \centerline{
      \includegraphics[width=0.5\linewidth]{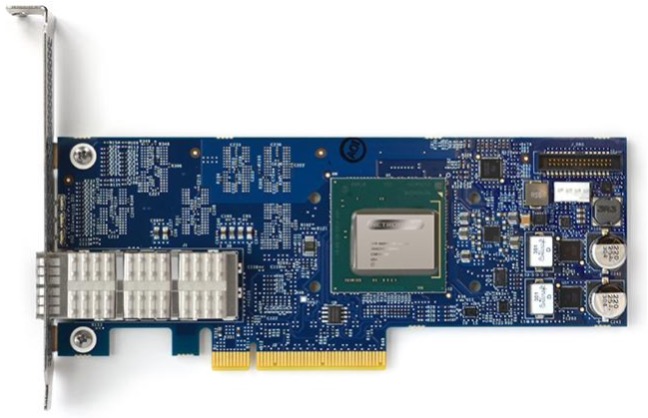}
      \includegraphics[width=0.5\linewidth]{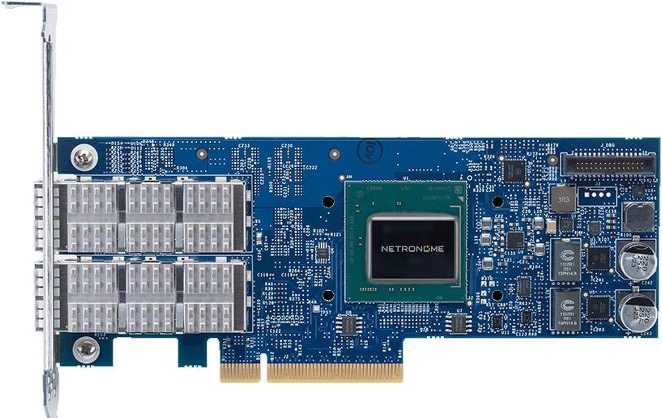}
    }
    \caption{Netronome Agilio CX single-port (left) and dual-port (right)
      SmartNICs used in this project for experimental prototype testing of
      enabling measurements from custom firmware-level modifications on NICs.}
      \label{fig:agilio_nics}
\end{figure*}

\begin{figure*}[!t]
    \centerline{
      \includegraphics[height=2.1in]{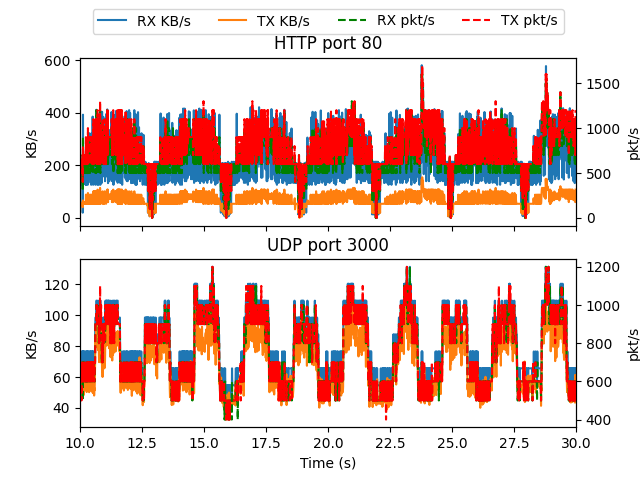}
          \includegraphics[height=2.1in]{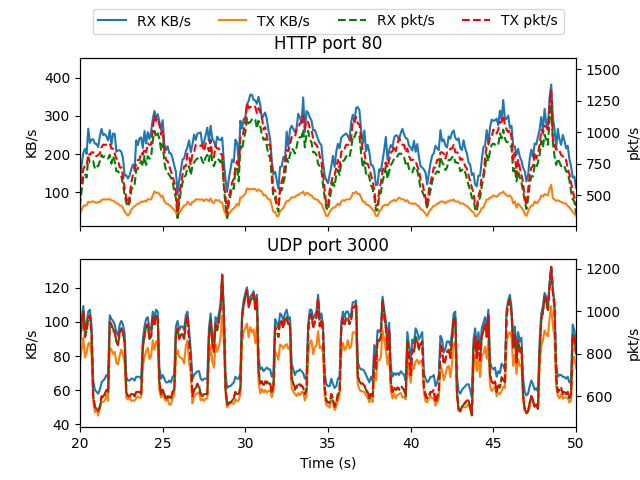}
      \includegraphics[height=2.3in]{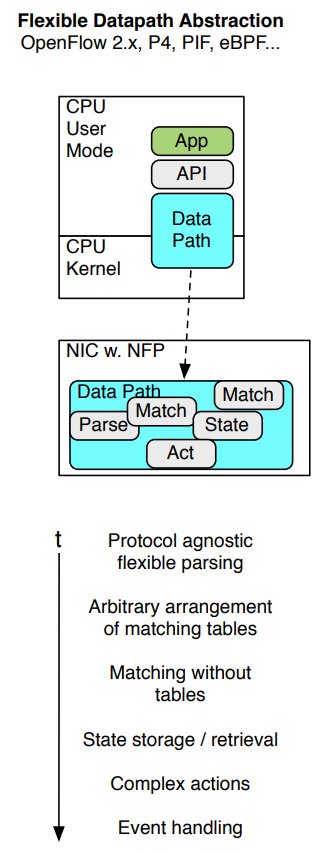}
    }
    \caption{Measurements of TCP and UDP packet counts and byte counts (transmitted, received) for different ports collected using modified firmware under different patterns of network traffic. Left: with sampling rate of 10ms; Middle: with sampling rate of 100ms; Right: Programming model for integration of firmware modifications in the CoreNIC structure to enable custom datapath with parsing, matching, and action functions.}
      \label{fig:nic_measurements}
\end{figure*}

A prototype implementation~\cite{udeshi2025tamperproof} was developed for collecting firmware-level
measurements from the Agilio SmartNIC. Through custom modifications of the NIC's firmware, measurements of packet counts and byte counts aggregated per port for both TCP and UDP in both transmitted and received directions were implemented.
The prototype for experimental testing uses Netronome Agilio CX Single-Port and Dual-Port 40 Gigabit Ethernet SmartNICs (Figure~\ref{fig:agilio_nics}) that are supported by the open-source CoreNIC firmware. The custom measurements are collected using firmware modifications of MicroC code and P4 logic on the NIC to enable custom datapath based on parsing, matching, and action functionalities using the manufacturer-provided firmware library. Sample measurements are shown in Figure~\ref{fig:nic_measurements} and include time series measurements of TCP and UDP packet and byte counts. These packet/byte counts are recorded per-port and for both transmitted and received packets. Measurements are collected using the custom-modified firmware under different patterns of network traffic. The time series measurements support configurable aggregation time intervals (measurements with sampling rates of 10ms and 100ms are shown in Figure~\ref{fig:nic_measurements}). 

\begin{figure*}[!t]
    \centerline{
        \includegraphics[width=0.5\linewidth,clip=true,trim=1in 0in 1in 0in]{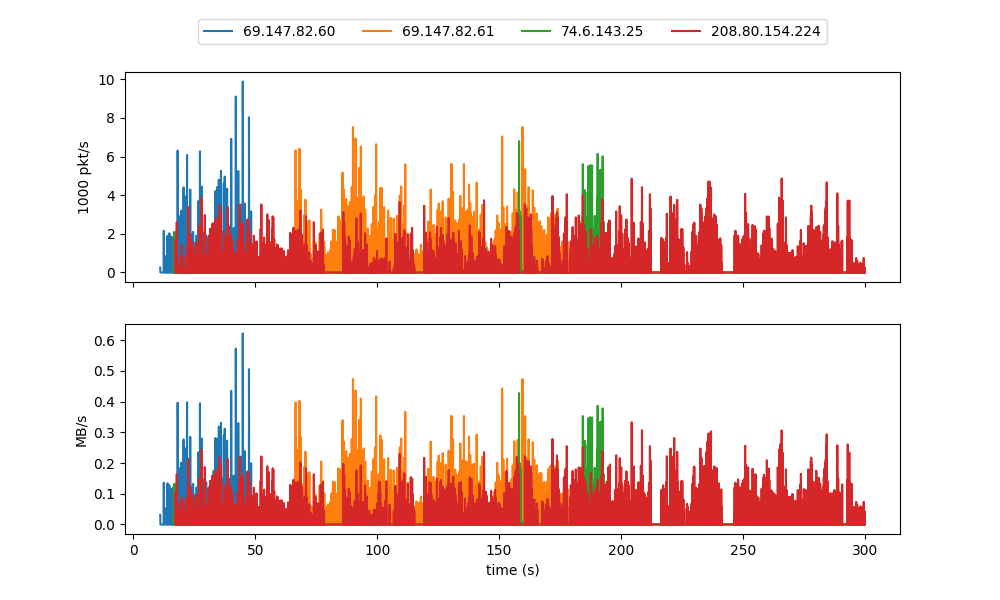}\ \
        \includegraphics[width=0.5\linewidth,clip=true,trim=1in 0in 1in 0in]{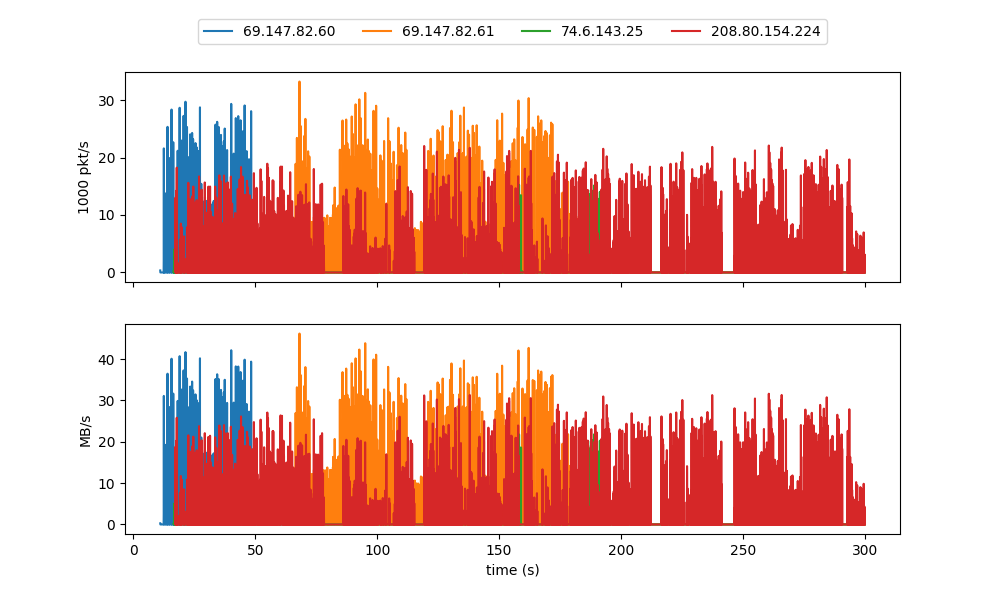}
    }
    \caption{Detection of IP addresses from within firmware of NIC. Time series of measurements of packet counts and byte counts (transmitted, received) for different ports and IP addresses collected using modified firmware. This data is from a test running recursive wget on yahoo.com (three IP addresses) and wikipedia.org (one IP address). Left: transmitted data; Right: received data.}
    \label{fig:NIC_data_IP}
\end{figure*}

The firmware-level measurements include detection of IP addresses and aggregation of observed traffic for each detected IP address.
IP addresses are detected for both transmitted and received TCP and UDP traffic from within the NIC firmware and time series of packet and byte counts are collected for each IP address (both incoming and outgoing traffic).  Sample time series measurements of packet and byte counts aggregated per IP address are shown in Figure~\ref{fig:NIC_data_IP} for a sampling rate of 20 ms.

{\em On-NIC encryption mechanism:}
Network traffic measurements (numbers of packets and numbers of bytes in both transmit and receive directions) for each detected IP address and port are collected in the NIC using custom-modified firmware. To enable the secure transiting of these measurements through untrusted intermediaries, a firmware-based mechanism was implemented for encrypting the collected measurements with timestamps onboard the NIC.  The on-NIC encryption of the measurements is implemented using Speck cipher with 64 bit blocks and 128 bit key. The Message Authentication Code (MAC) for the NIC-generated payload is constructed by hashing the encrypted payload using HalfSipHash using a 64 bit key.  The sending of measurements (stored in a table in on-NIC memory) is triggered using an Ethernet control packet of a custom type with a request (table index). The control packet can be sent from the host computer or a remote computer, thereby also enabling measurements to be directly transmitted either to the host or a remote computer completely independently of the host computer.  Furthermore, timestamps are integrated into the encrypted measurements to prevent replay attacks. Since the measurements are recorded directly on the NIC using custom-modified firmware, the collected measurements can be used to detect malware that hide network activity from the host using a rootkit (e.g., elf.reptile). For such malware, comparison of the NIC-reported network traffic with observations on the host computer enables detection of the malware.
To verify that the on-NIC encryption does not impose significant overhead
leading to reduction in network communication speeds, we performed network speed
tests (both in terms of packets per second and bytes per second) without and
with encryption enabled. It was found that the encryption overhead is negligible and does not significantly impact the network communication speeds.

\subsection{GPU}
\begin{figure}[!h]
    \centerline{
      \includegraphics[width=0.8\linewidth]{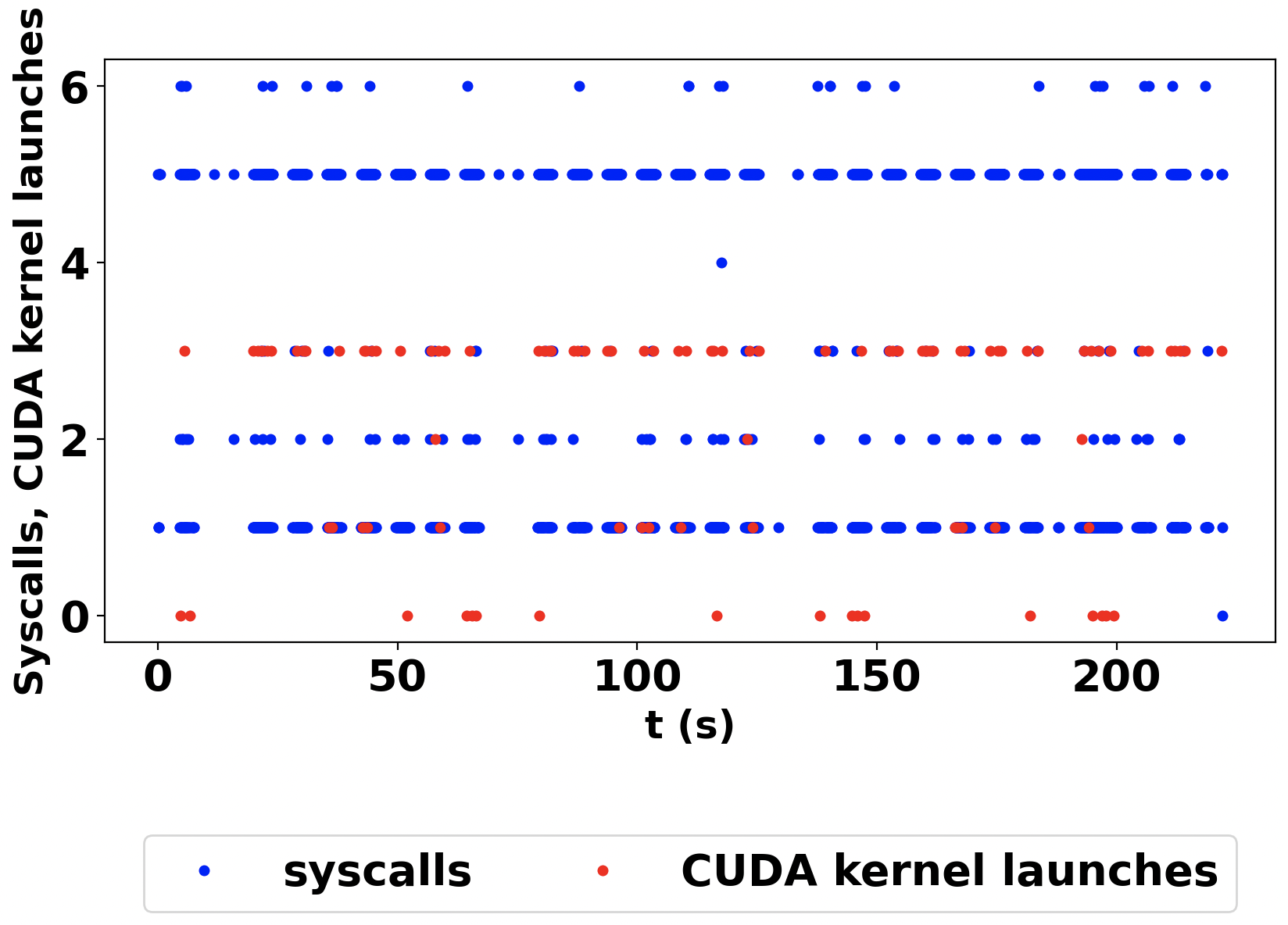}
    }
    \caption{Time series measurements of system calls (blue dots) and CUDA kernel launches (red dots). The Y-axis corresponds to different system calls and CUDA kernel launches detected in this test as follows: system calls -- [`write', `futex', `sendmsg', `restart\_syscall', `poll', `read', `ioctl']; CUDA kernels -- [`reduce\_kernel', `vectorized\_elementwise\_kernel', `memcpy32\_post', `elementwise\_kernel'].}
      \label{fig:GPU_syscalls_kernels}
  \end{figure}

\begin{figure*}[!t]
    \centerline{
      \includegraphics[height=1.7in]{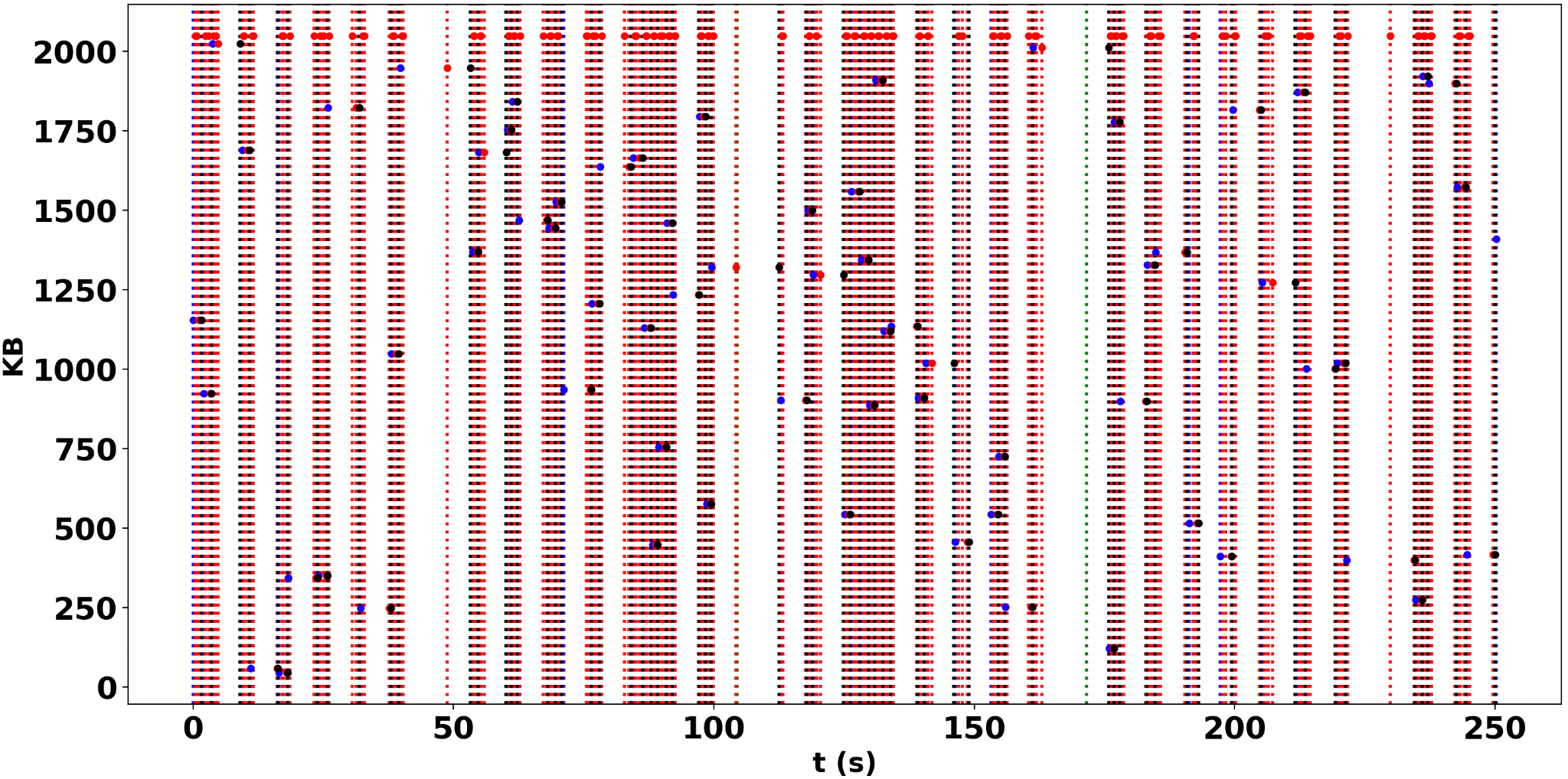}\ \
      \includegraphics[height=1.7in]{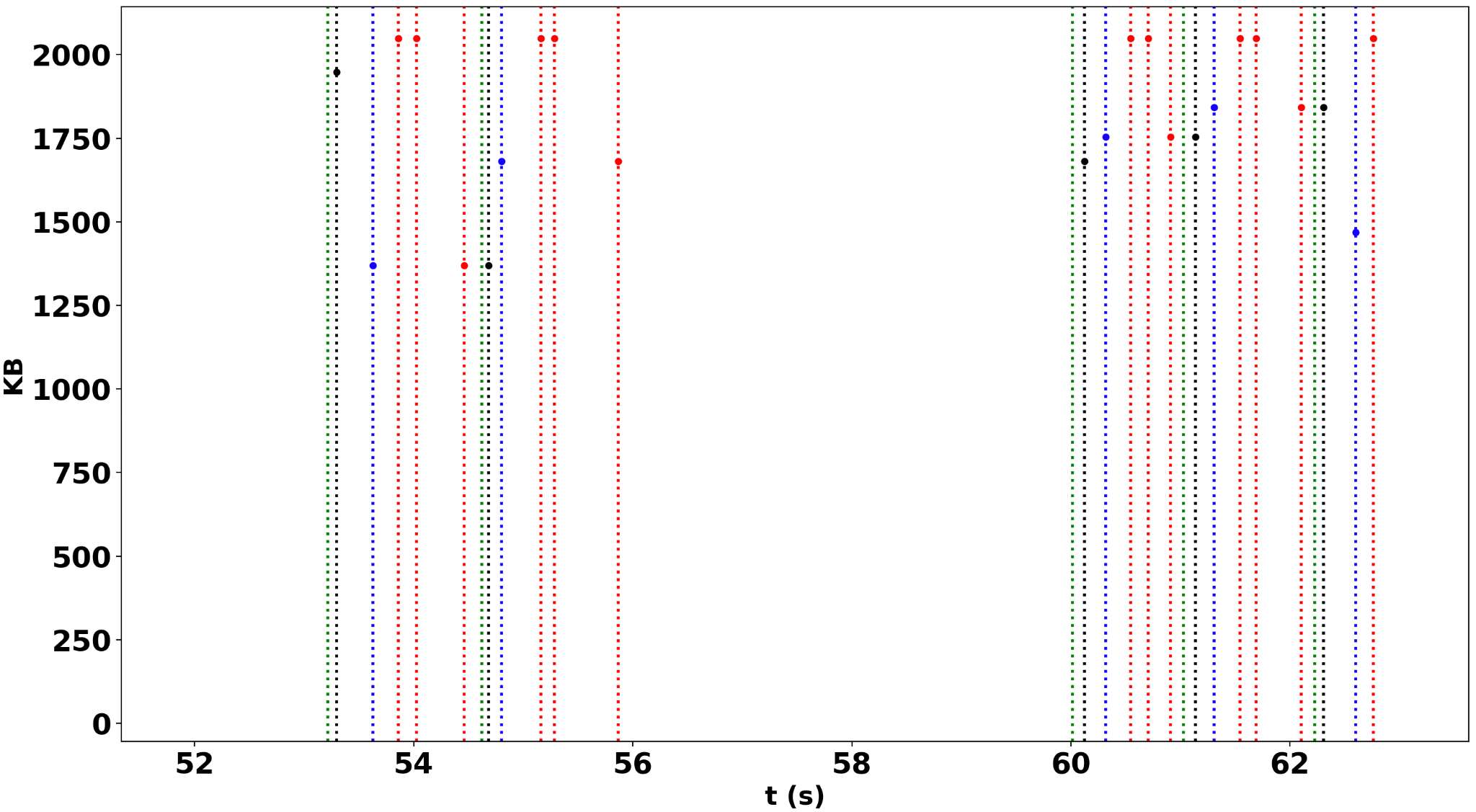}
    }
    \caption{Time series measurements of GPU memory events. Blue: GPU memory allocations; green: GPU memory frees; red: CPU-to-GPU memory copies; black: GPU-to-CPU memory copies. Right-side plot is a zoomed-in view of the left-side plot. Y-axis indicates memory sizes allocated/copied.
    }
      \label{fig:GPU_memory_events}
  \end{figure*}

\begin{figure}[!h]
    \centerline{
      \includegraphics[width=0.95\linewidth,clip=true,trim=0in 0in 0in 0in]{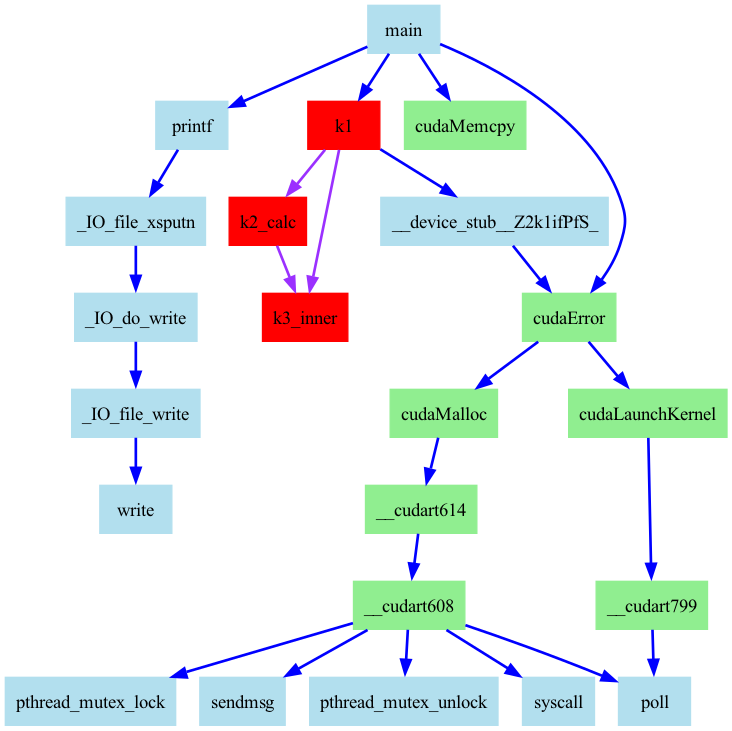}
    }
    \caption{Reconstructed call graph including CPU and GPU activity. Red: CUDA kernels; green: CUDA library functions; blue: other CPU functions.}
      \label{fig:GPU_call_graph}
  \end{figure}

A prototype implementation was developed for collection of measurements of GPU activity and CPU-GPU interaction.
    Timestamps of CUDA kernel launches and system calls are simultaneously tracked. On detected system calls, backtraces are collected and system call arguments are captured.
    Time series measurements are simultaneously collected for all processes using the GPU in the system. Sample time series measurements are shown in Figure~\ref{fig:GPU_syscalls_kernels} for a process using the PyTorch python library.
The initial prototype implementation for collecting measurements from the GPU was refined to expand the set of measurements collected of the GPU activity and CPU-GPU interaction. The measurements obtained include simultaneous tracking of time series of CUDA kernel launches and system calls, interrelationships between kernels (kernel launches by other kernels), backtraces, reconstruction of call graph including GPU and CPU activity, GPU thread events (thread synchronize, thread exit), and tracking of GPU memory events (allocations with recording of memory sizes allocated, memory frees, memory copies including CPU-to-GPU and GPU-to-CPU) with timestamps. Sample time series measurements of GPU memory events is shown in Figure~\ref{fig:GPU_memory_events}. A sample reconstructed call graph including both GPU and CPU activity is shown in  Figure~\ref{fig:GPU_call_graph}.    

\subsection{Keyboard}
  Firmware modifications of a keyboard were implemented to enable collection of fine-grained measurements of keyboard activity as well as software-driven loopbacks. 
The prototype implementation for experimental testing was based on firmware modifications of the open-source QMK firmware, which supports several keyboards such as the Moonlander and ErgoDox EZ.  The experimental testing was performed primarily on the Moonlander keyboard, which has a STM32F303 processor running the ChibiOS real-time operating system. The custom-modified firmware is cross-compiled to ARM using gcc and the bin file is loaded to the keyboard.

\begin{figure}[!h]
\centerline{
\includegraphics[height=0.9in]{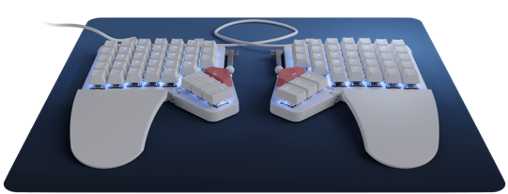}
\includegraphics[height=0.9in]{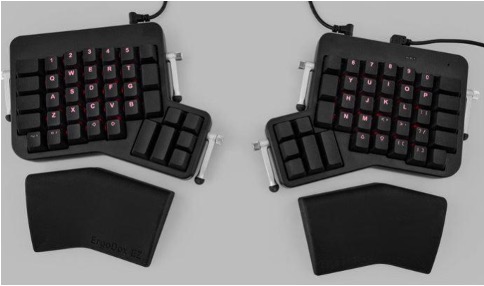}
}
\caption{Moonlander (left) and ErgoDox EZ (right) keyboards used in this
work for experimental prototype testing of enabling measurements from keyboards.}
\end{figure}

\begin{figure}[!h]
\centerline{
\includegraphics[width=0.5\linewidth]{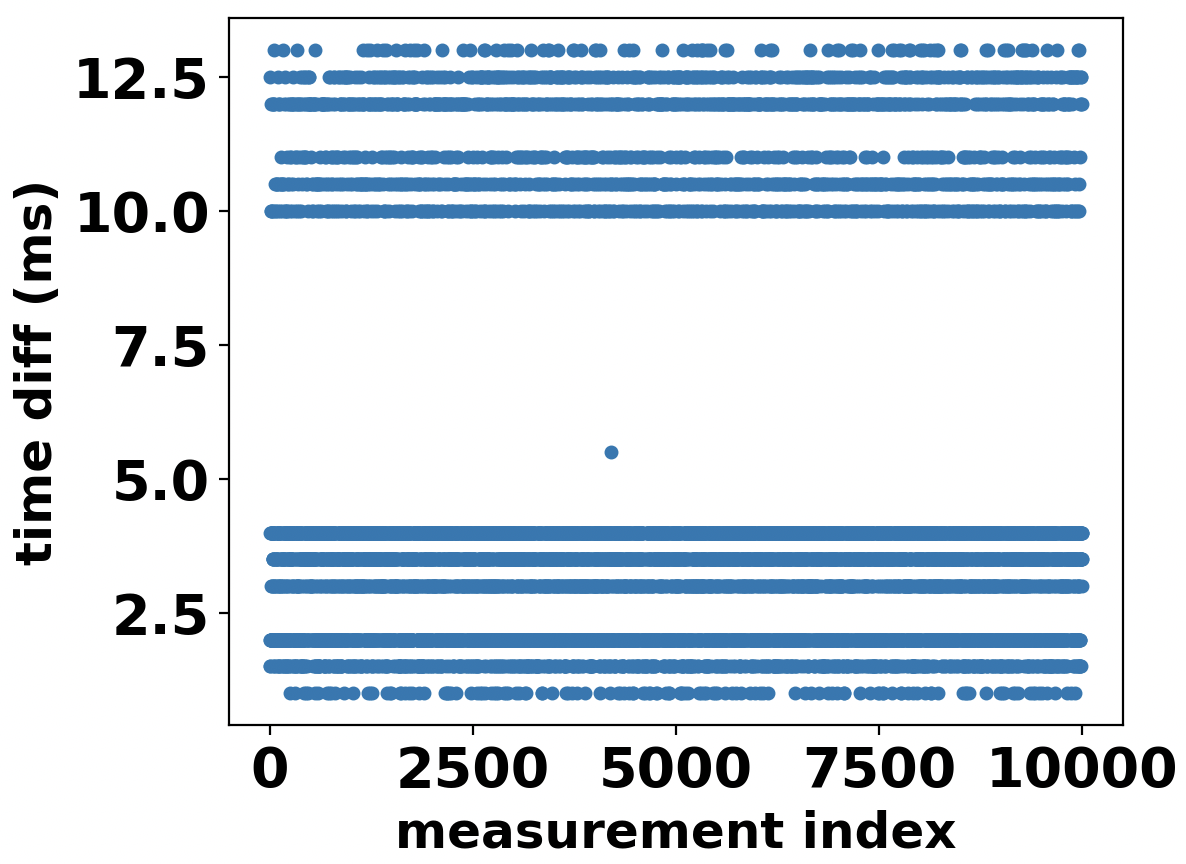}\ \
\includegraphics[width=0.5\linewidth]{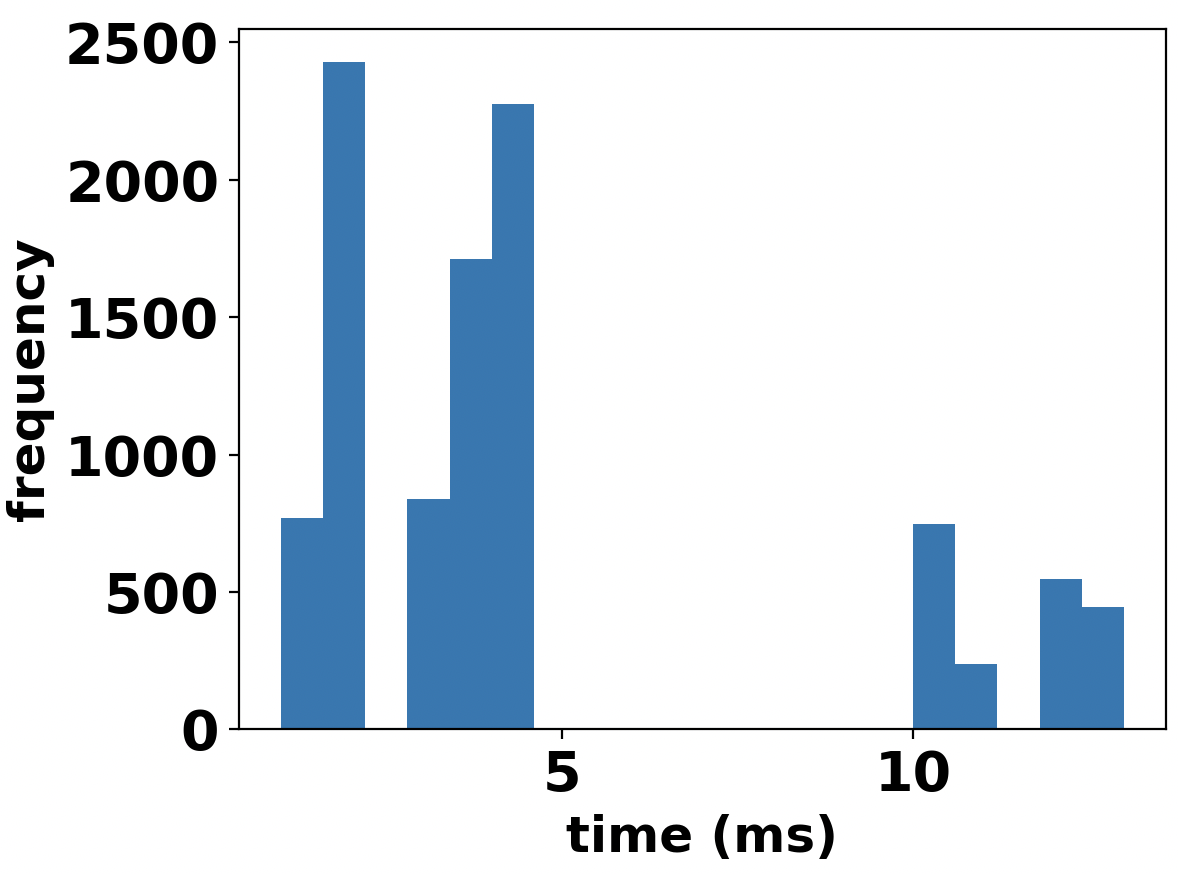}
}
\caption{Measurements of loopback times (with echo on keyboard and counter-based traffic generation from computer to bidirectional Raw HID channel). Left: measurement time series; Right: histogram of measurements. }
\label{fig:keyboard_loopback_times}
\end{figure}

\begin{figure}[!h]
\centerline{
\includegraphics[width=0.5\linewidth]{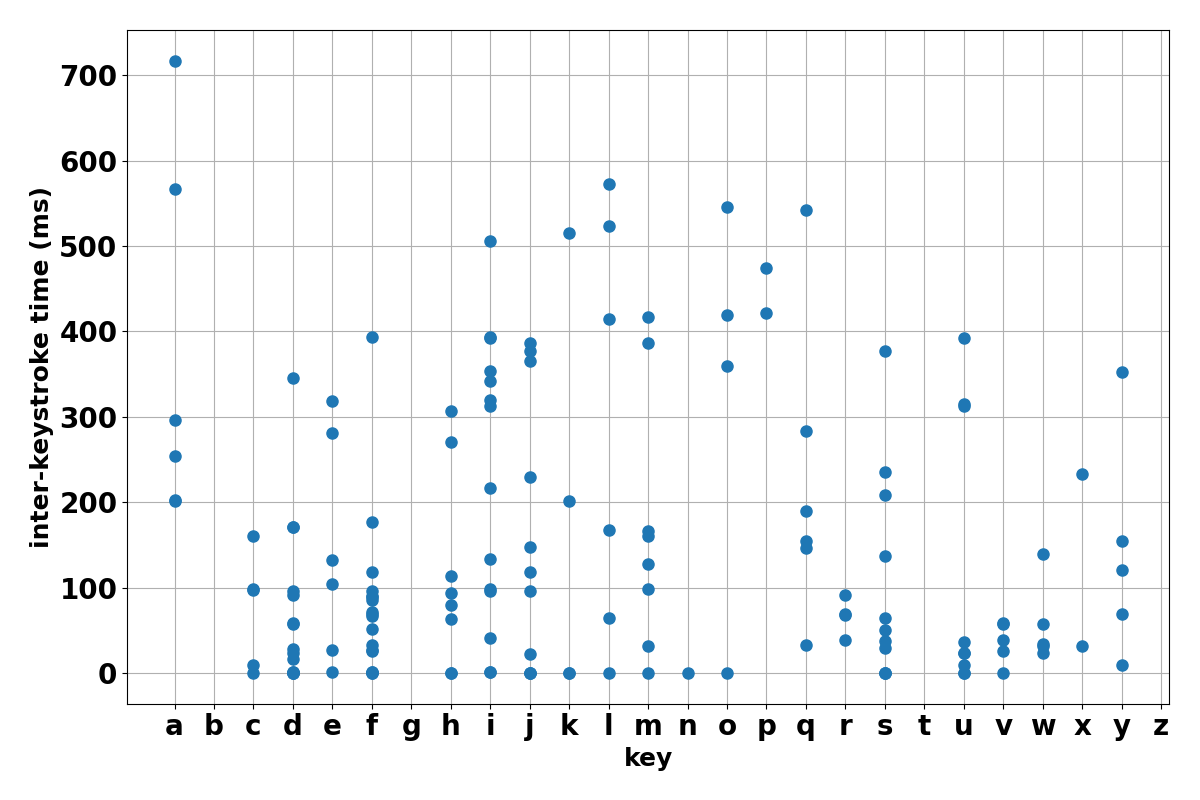}\ \
\includegraphics[width=0.5\linewidth]{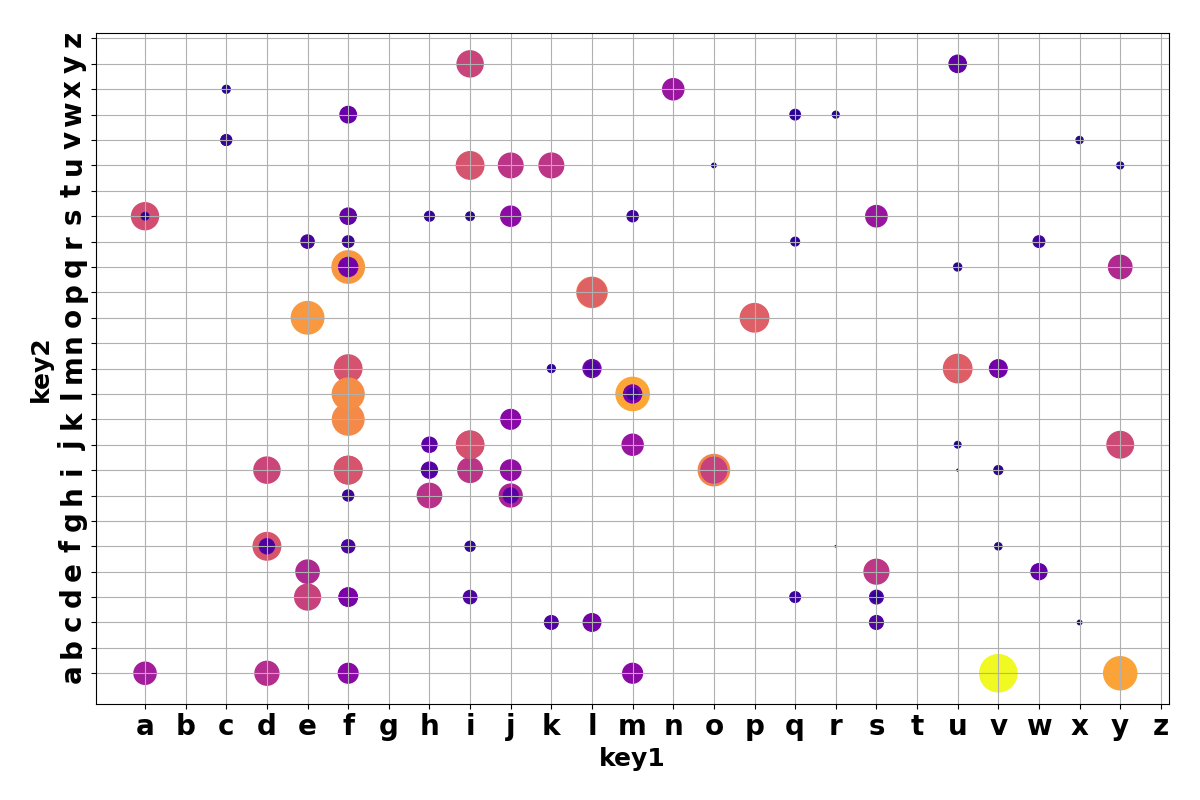}
}
\caption{ Left: Measurements of inter-keystroke timing (time intervals between every pair of keystrokes) shown as a time series; Right: Inter-keystroke times shown as a function of previous key (key1) and current key (key2) with dot sizes and colors indicating inter-keystroke timing magnitudes.  %
}
\label{fig:keyboard_inter_keystrokes}
\end{figure}

\begin{figure}[!h]
\centerline{
\includegraphics[height=0.9in]{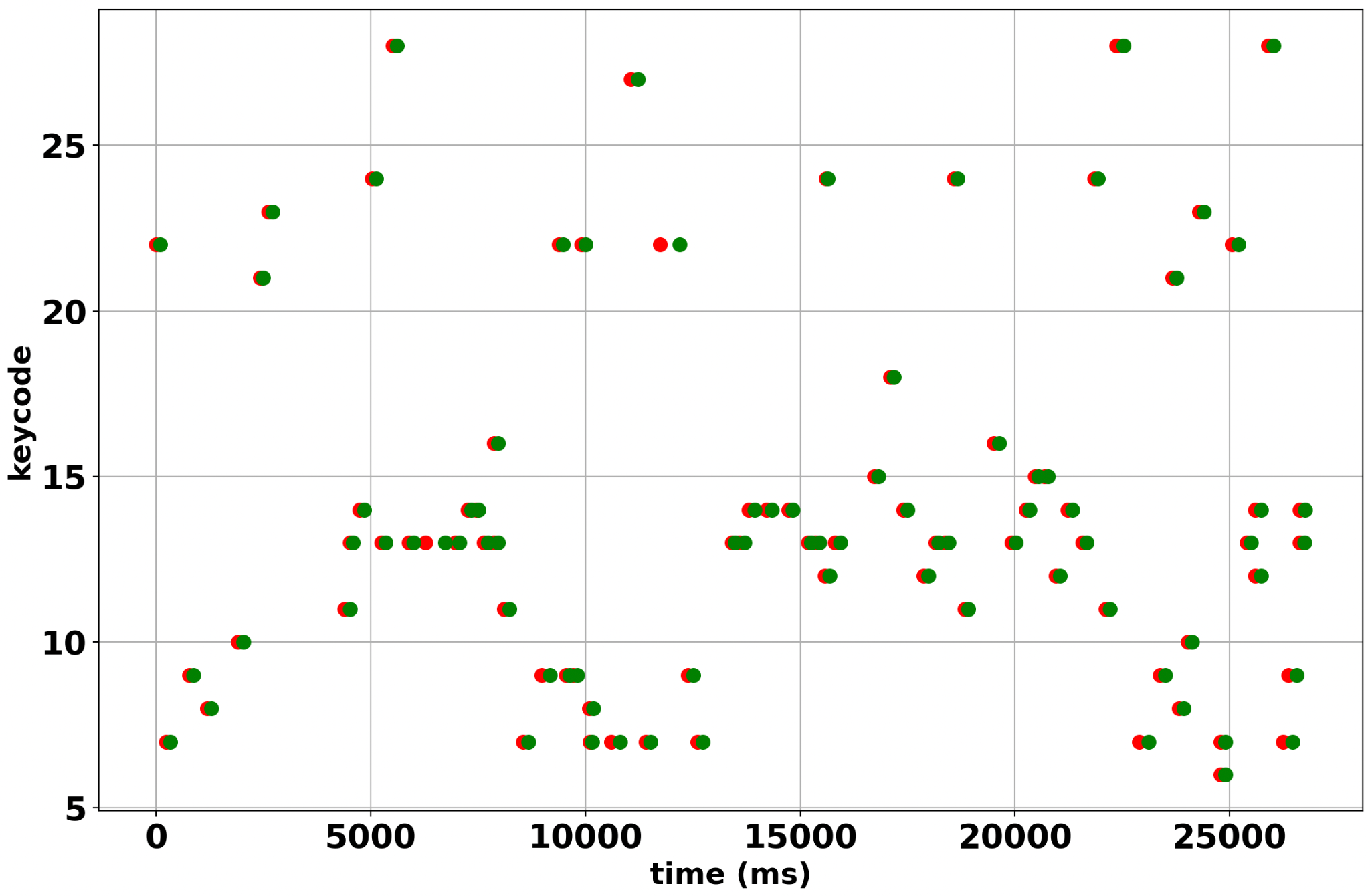}\ \
\includegraphics[height=0.9in]{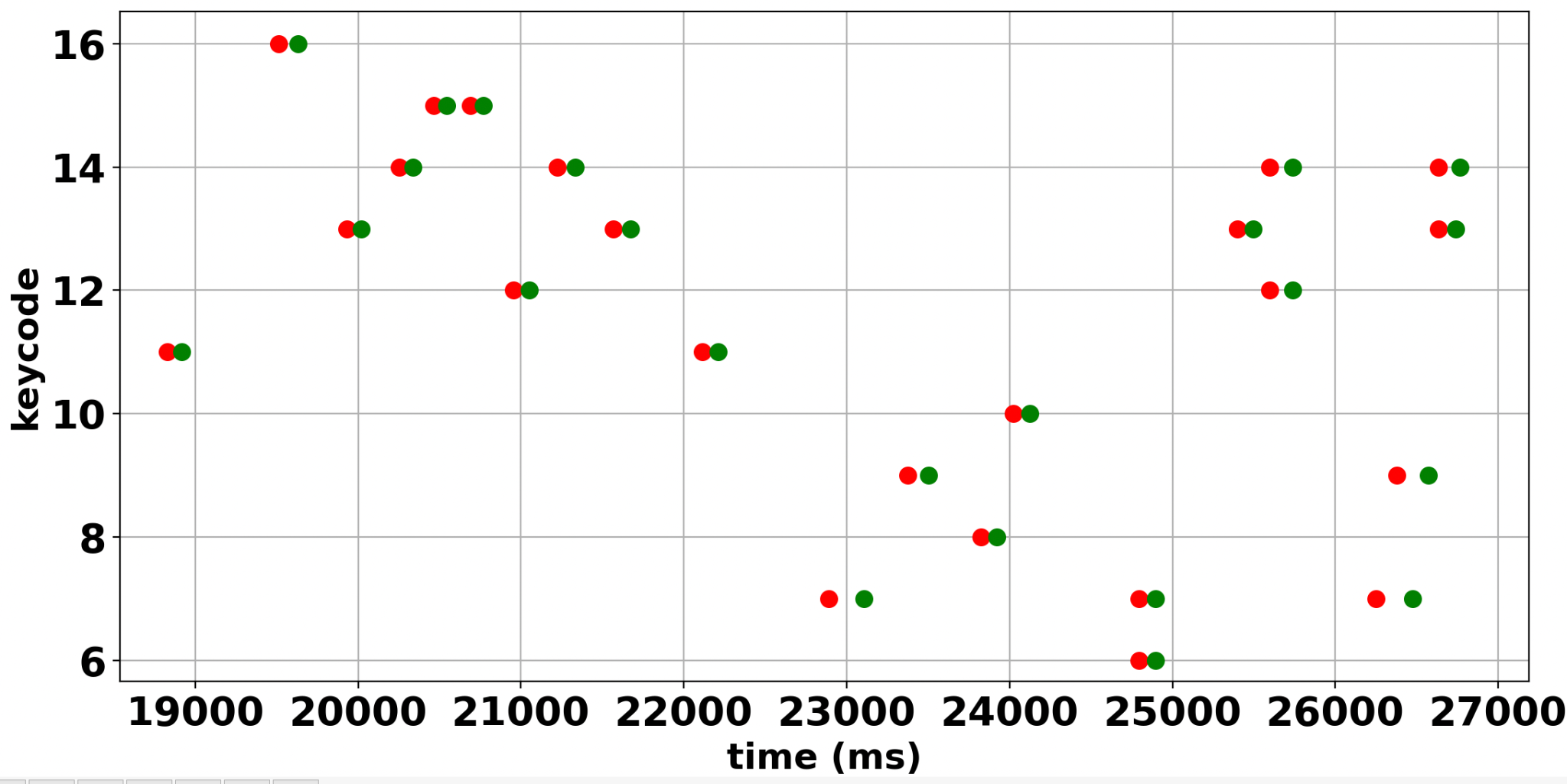}
}
\caption{Measurements of key press (red dots) and release (green dots) times. Right-side picture is a zoomed-in view.}
\label{fig:keyboard_press_release}
\end{figure}

\begin{figure}[!h]
\centerline{
\includegraphics[width=0.5\linewidth]{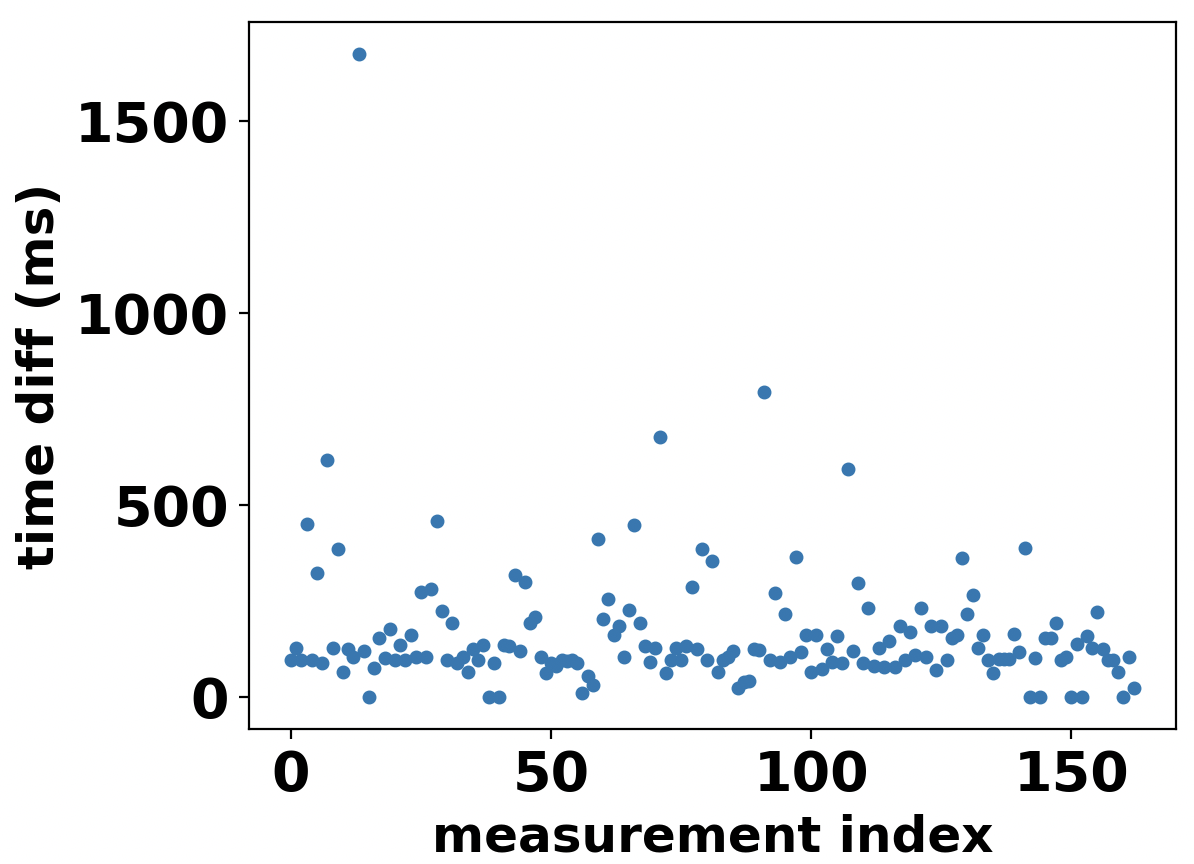}\ \
\includegraphics[width=0.5\linewidth]{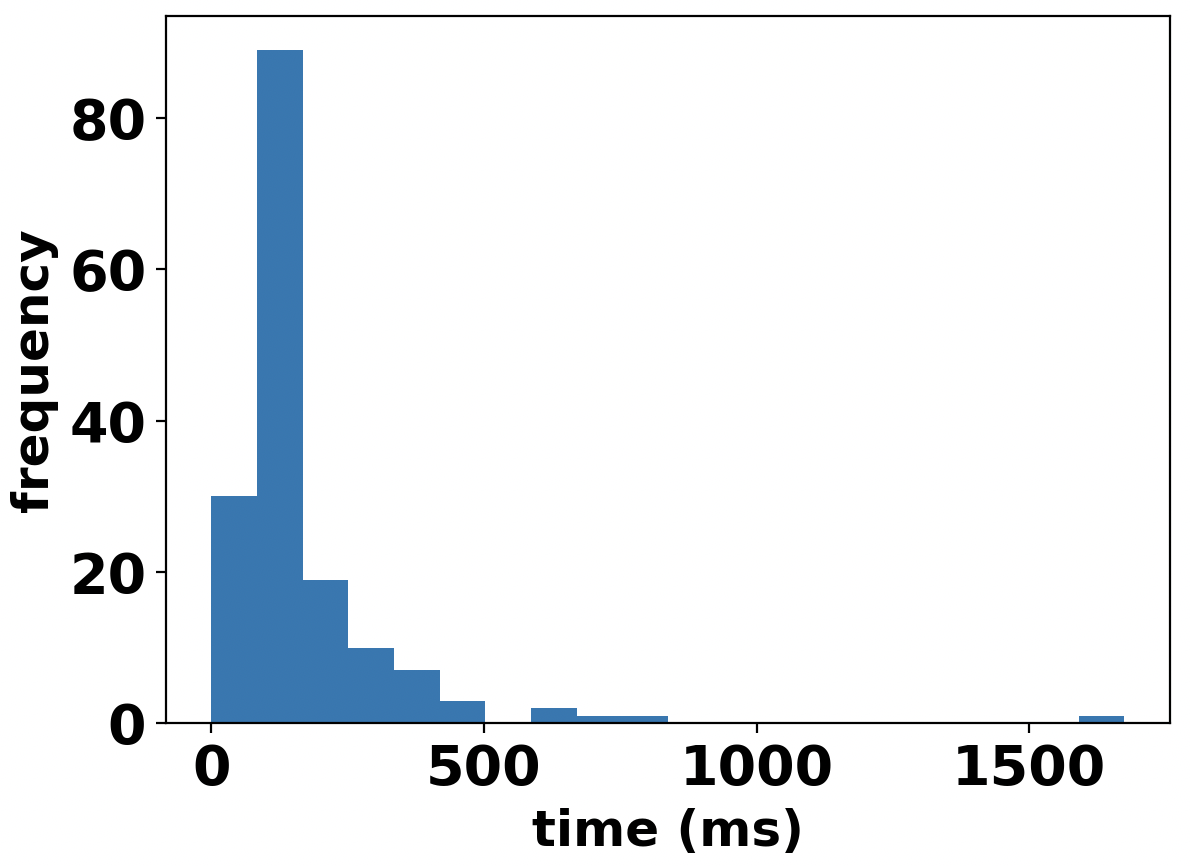}
}
\caption{Measurements of key dwell times (time from key press to key release). Left: measurement time series; Right: histogram of measurements.}
\label{fig:keyboard_dwell_times}
\end{figure}

\begin{figure}[!h]
\centerline{
\includegraphics[width=0.5\linewidth]{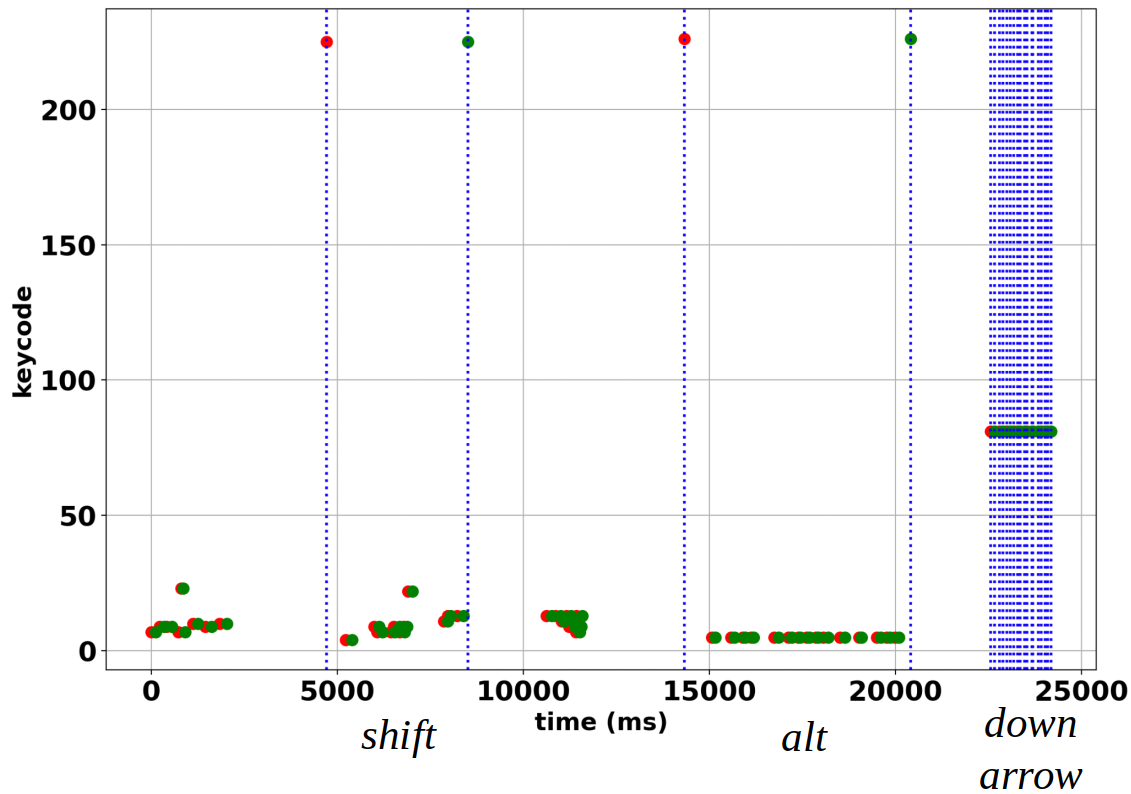}\ \
\includegraphics[width=0.5\linewidth]{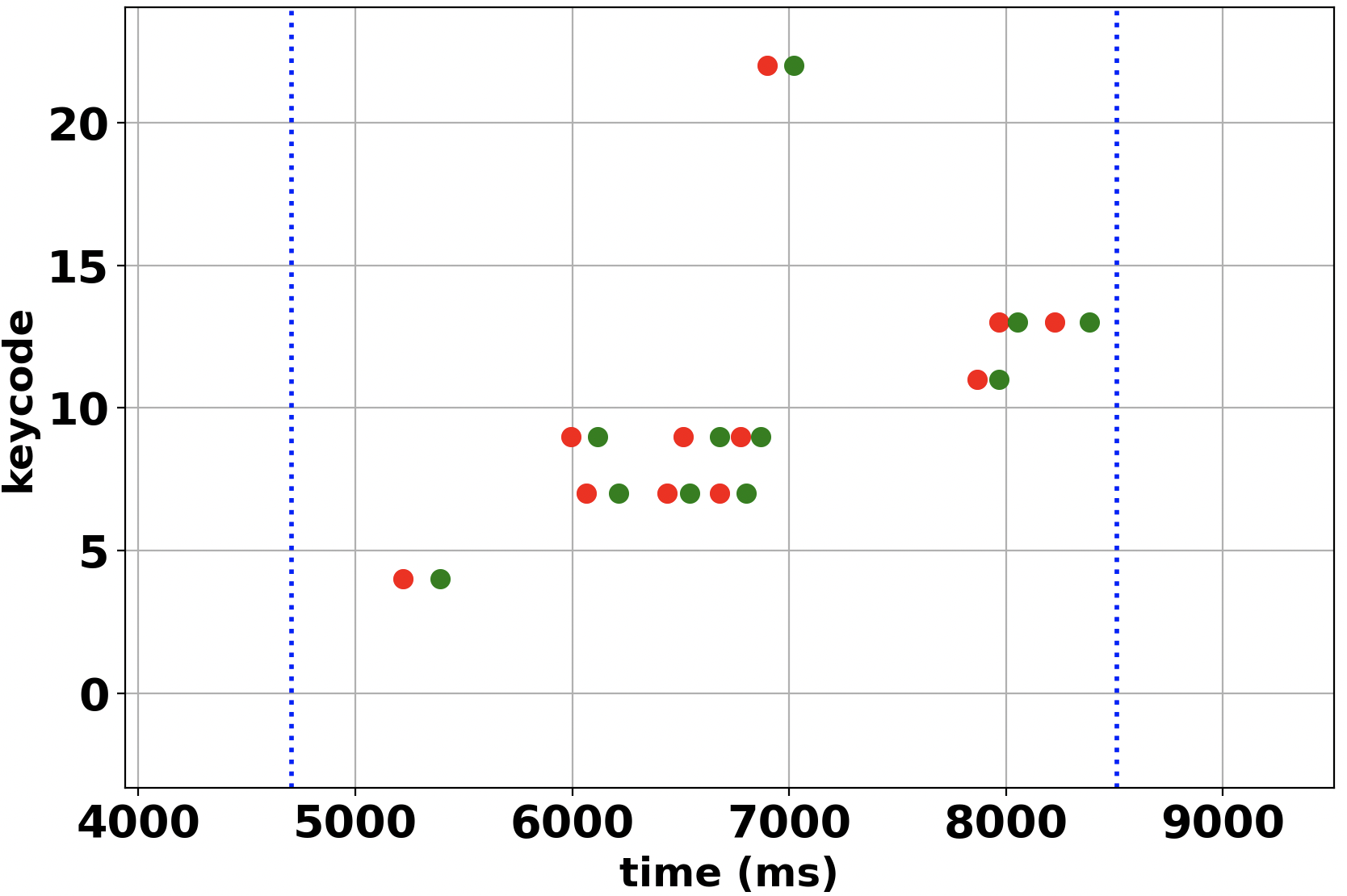}
}
\caption{Measurements of key press (red dots) and release (green dots) times including modifier keys. Presses and releases of non-[a-z,0-9] are shown as vertical dashed lines. Time intervals of shift-pressed, alt-pressed, and down arrow key presses and releases are shown. Right-side picture is a zoomed-in view.}
\label{fig:keyboard_modifier_keys}
\end{figure}

\begin{figure}[!h]
\centerline{
\includegraphics[width=0.47\linewidth]{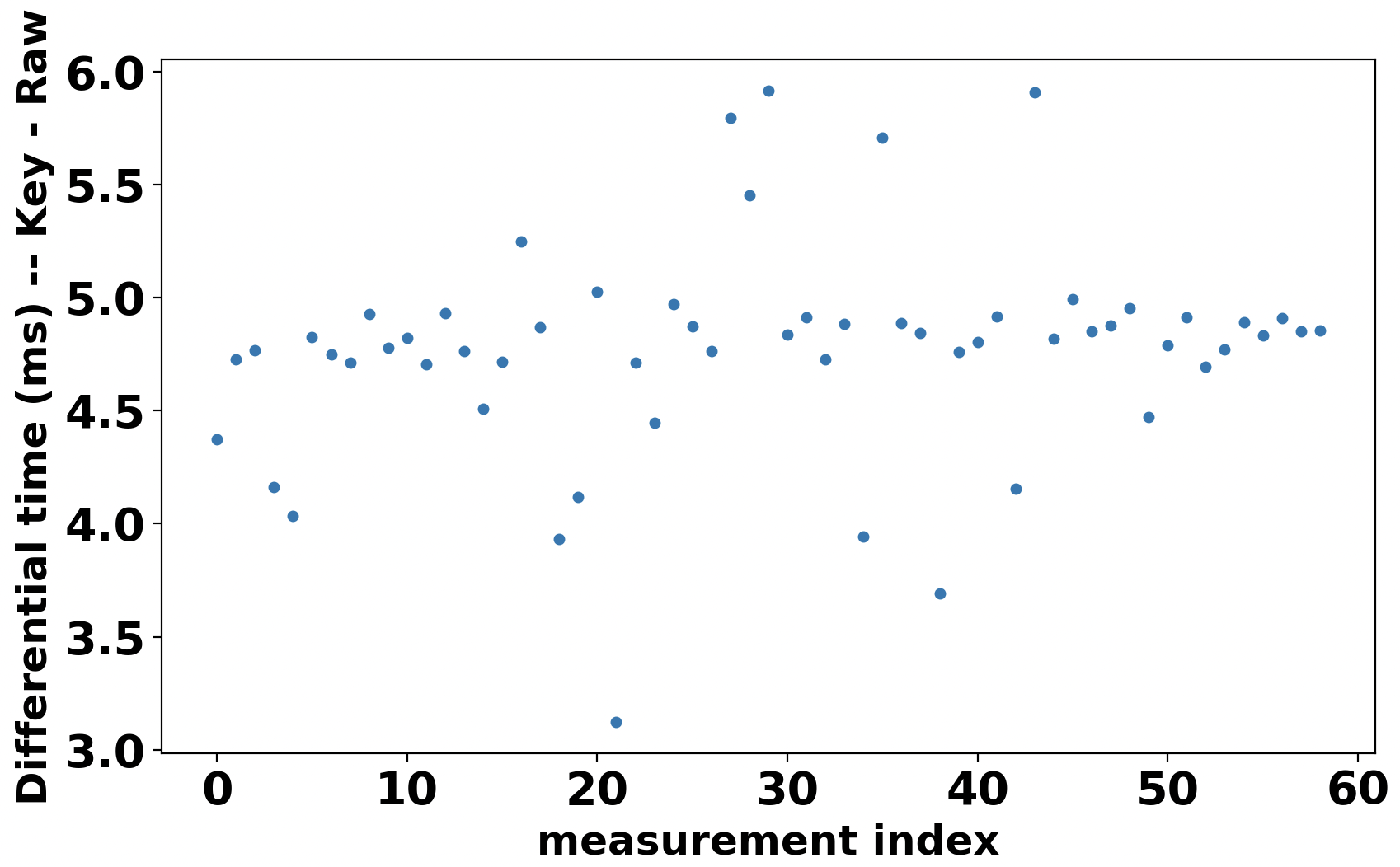}\ \
\includegraphics[width=0.48\linewidth]{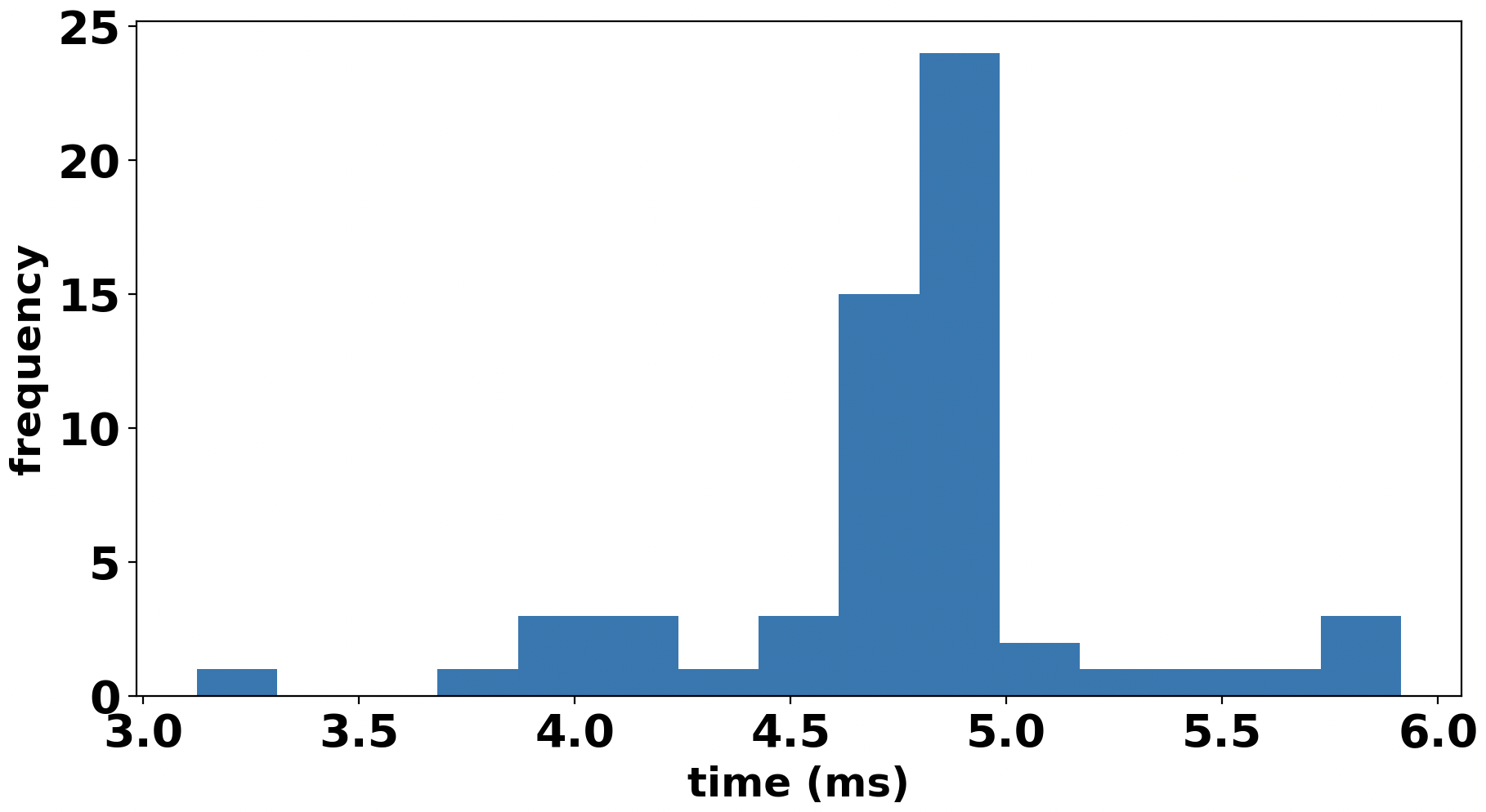}
}
\caption{Measurements of differential timing between key press event communicated via separate Raw HID channel and via standard keyboard interface. Left: measurement time series; Right: histogram.
}
\label{fig:keyboard_differential_timing}
\end{figure}

Several specific types of measurements were implemented as discussed below.
\begin{itemize}
    \item {\em Loopback timing measurements via Raw HID:} For exfiltration of measurements from the modified firmware, the ``Raw HID (Human Interface Device)'' mode was used that enables bidirectional communication between the computer and the keyboard. The USB-based raw mode is a separate binary bidirectional communication channel that operates independently of the normal keyboard behavior. As a representative example of custom measurements, 
    a dynamic loopback was implemented using the bidirectional communication between the computer and the keyboard through a firmware modification that echoes bytes received via Raw HID. In the loopback implementation, dynamic delays can be injected both on the computer and keyboard ends as a moving target challenge to an adversary. Sample loopback measurements are shown in Figure~\ref{fig:keyboard_loopback_times}.
    \item {\em Firmware-level recording of keystroke activity:}
    Custom measurements were implemented to capture user typing statistics including timing of keystrokes and inter-keystroke times. Sample measurements are shown in Figure~\ref{fig:keyboard_inter_keystrokes}. The timing measurements are obtained from the real-time clock on the STM32F303 processor in the keyboard. These measurements enable capturing a signature of user typing patterns that could be used to, for example, detect unauthorized access and keyboard spoofing.
     In addition, the measurements were expanded to include separate timing measurements for key presses and releases (Figure~\ref{fig:keyboard_press_release}) and thereby key dwell times (Figure~\ref{fig:keyboard_dwell_times}).  
      The time intervals for transitions between different pairs of keys and the per-key dwell times provide distinctive markers of user typing patterns. The key press and release measurements also support modifier keys and special characters (e.g., non-[a-z,0-9]) and therefore enable detecting simultaneous holding of keys (e.g., holding down shift while typing other keys) as seen in Figure~\ref{fig:keyboard_modifier_keys}. The measurements are in terms of keycodes and supports non-English languages as well.
      Measurements are transmitted using the Raw HID mode. A streaming mode was implemented in the Raw HID mode with increased communication bandwidth and reduced latencies, enabling real-time visibility into operations on keyboard.
    \item {\em Differential timing between Raw HID and keyboard interface:}
Measurements of differential timings between key press events communicated via the separate Raw HID channel and via the standard keyboard interface were implemented. Being a separate USB-based binary bidirectional communication channel that operates independent of keyboard behavior,
the Raw HID channel is faster and ``out of band'' while the standard keyboard interface goes through OS libraries for keystroke handling and dispatch to applications. Hence, when a key is pressed, the notification of the event sent from our modified firmware is received first over Raw HID and after a short delay, over the standard keyboard interface. The differential time between the receiving of the event over Raw HID and over the standard keyboard interface is indicative of the latencies in the OS libraries and other software layers through which the keyboard events are processed. 
Therefore, the differential timing measurements can help in illuminating latency variations due to changes in libraries through which the keyboard events traverse, e.g., keyloggers.   
Sample measurements of differential timing are shown in Figure~\ref{fig:keyboard_differential_timing}. 
\end{itemize}

\subsection{CPU HPCs and Power}
\begin{figure}[!h]
    \centerline{\includegraphics[width=0.95\columnwidth]{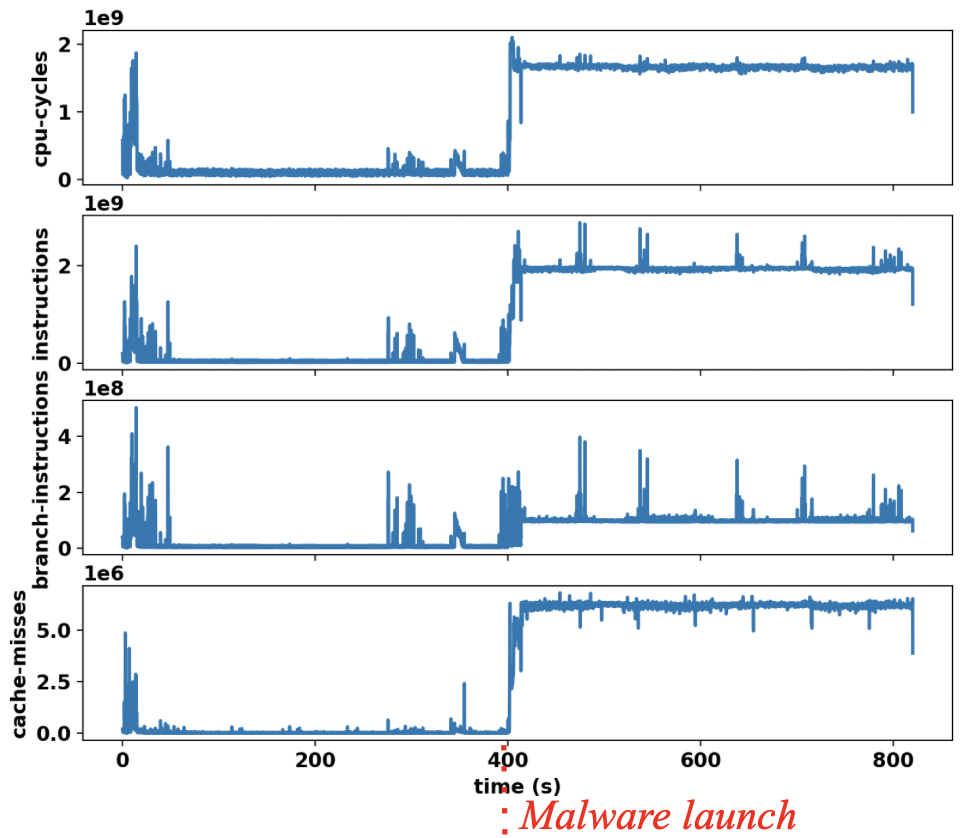}
  }
  \vspace*{-0.1in}
  \caption{A sample of CPU HPC time series measurements collected for elf.capoae malware
    executable from malpedia.}
  \label{fig:capoae_hpc}
  \vspace*{-0.1in}
\end{figure}

In parallel with the collection of the subcomponent measurements outlined in
this section,
the integrated measurement framework implemented in this project collects
multiple HPC measurements (including numbers of CPU cycles, instructions, branch
instructions, branch misses, cache misses, cache references, memory loads, and
memory stores) as well as CPU power usage measurements as time series. These
time series measurements are obtained using the {\em perf} command-line tool on
the Linux host. While the focus of the proposed approach is on subcomponent measurements
independent of the main processor, the inclusion of processor-level HPC~\cite{krishnamurthy2020anomaly} and
power usage measurements provides an auxiliary information source that, in
combination with the other subcomponent measurements, can enable flagging of
discrepancies among the measurements and thereby a more comprehensive view of
the system operation enabling more robust anomaly detection. A sample of HPC
measurements collected for the elf.capoae malware sample from malpedia are shown
in Figure~\ref{fig:capoae_hpc}. The increased system activity after the launch
of the malware sample (at $t=400$s) are captured in the HPC measurements as
shown in Figure~\ref{fig:capoae_hpc}.
As shown in Figure~\ref{fig:malware_testing_timeline}, a kill command is issued
at $t = 700 s$ to terminate the original process that was started for the
malware sample. Apparently, this malware sample (similar to many other malware
samples as well) starts other processes and/or infiltrates the kernel via a
rootkit so that it continues operation even after the kill command is issued to
terminate the original launched malware process. The data collection is
continued until $t = 820 s$.

\section{Subcomponent-Level Data Collection and Analysis}
\label{sec:analysis}
Using the experimental testbed and dataset collection framework described in
Section~\ref{sec:testbed}, subcomponent time series datasets were collected on
the integrated testbed for a large variety of malware samples.
To ensure that the data collection between different samples is independent, the
VM-based approach discussed in Section~\ref{sec:testbed} was used wherein the VM
image is restored between successive runs. 
The time settings for the data collection are uniform for each malware sample as
shown in Figure~\ref{fig:malware_testing_timeline}.
For all the malware samples, the data collection is started when the VM is
launched. The malware sample is launched at $t=400 s$. At $700 s$, an attempt is
made to kill the original malware process that was started. For several malware,
the original malware process could start other processes and/or infiltrate the
kernel via a rootkit in which case the operation would continue even after the
kill command is issued to terminate the original launched malware process. The
data collection is continued until $t=820 s$.
The measurements for each
subcomponent are collected as time series data into csv files. The timestamps
are synchronized across all the subcomponents.

\begin{figure}[!h]
  \centerline{
    \includegraphics[width=0.95\linewidth]{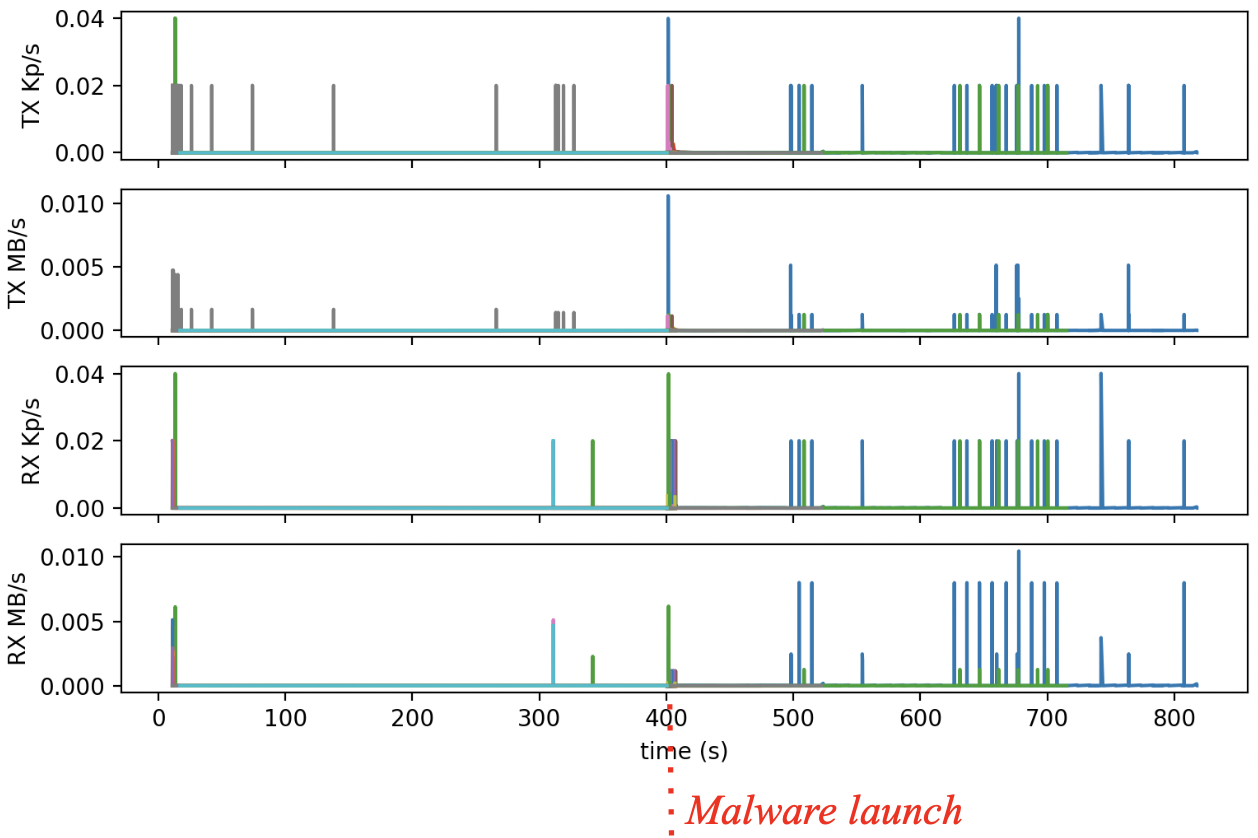}}
    \vspace*{-0.1in}
    \caption{NIC time series measurements collected for elf.capoae malware sample.}
  \label{fig:capoae_nic}
  \vspace*{-0.1in}
\end{figure}

\begin{figure}[!h]
\centerline{
  \includegraphics[width=0.9\linewidth]{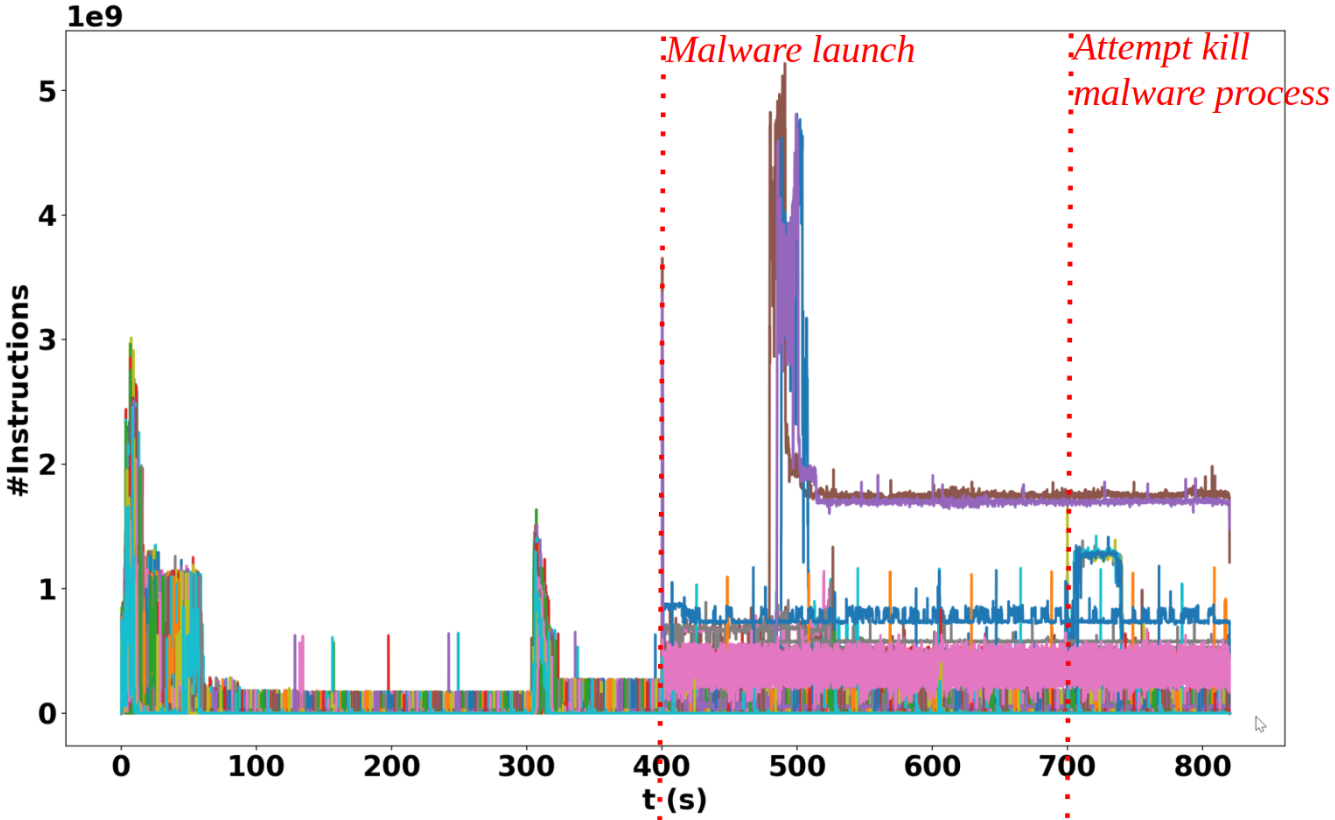}
}
\centerline{
  \includegraphics[width=0.9\linewidth]{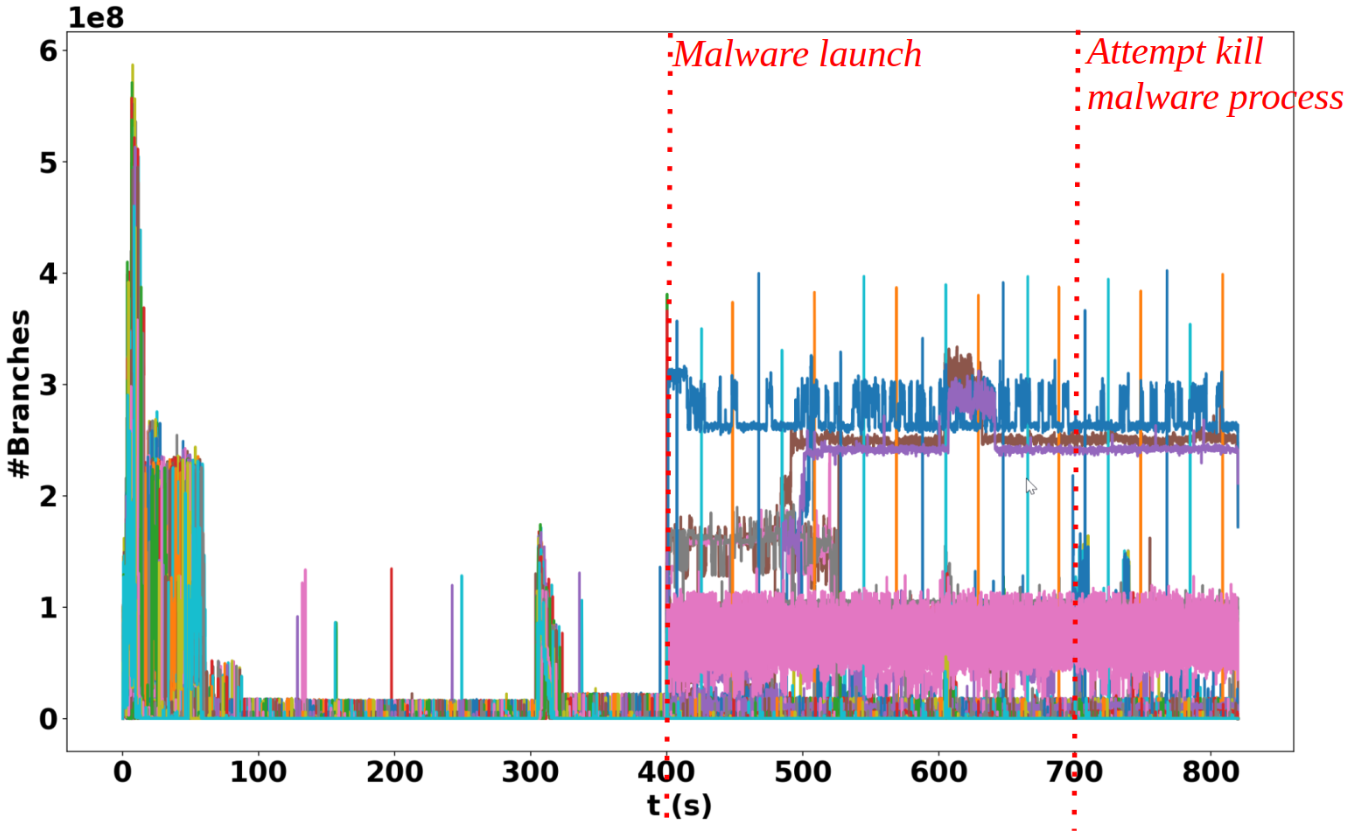}
}
\centerline{
  \includegraphics[width=0.9\linewidth]{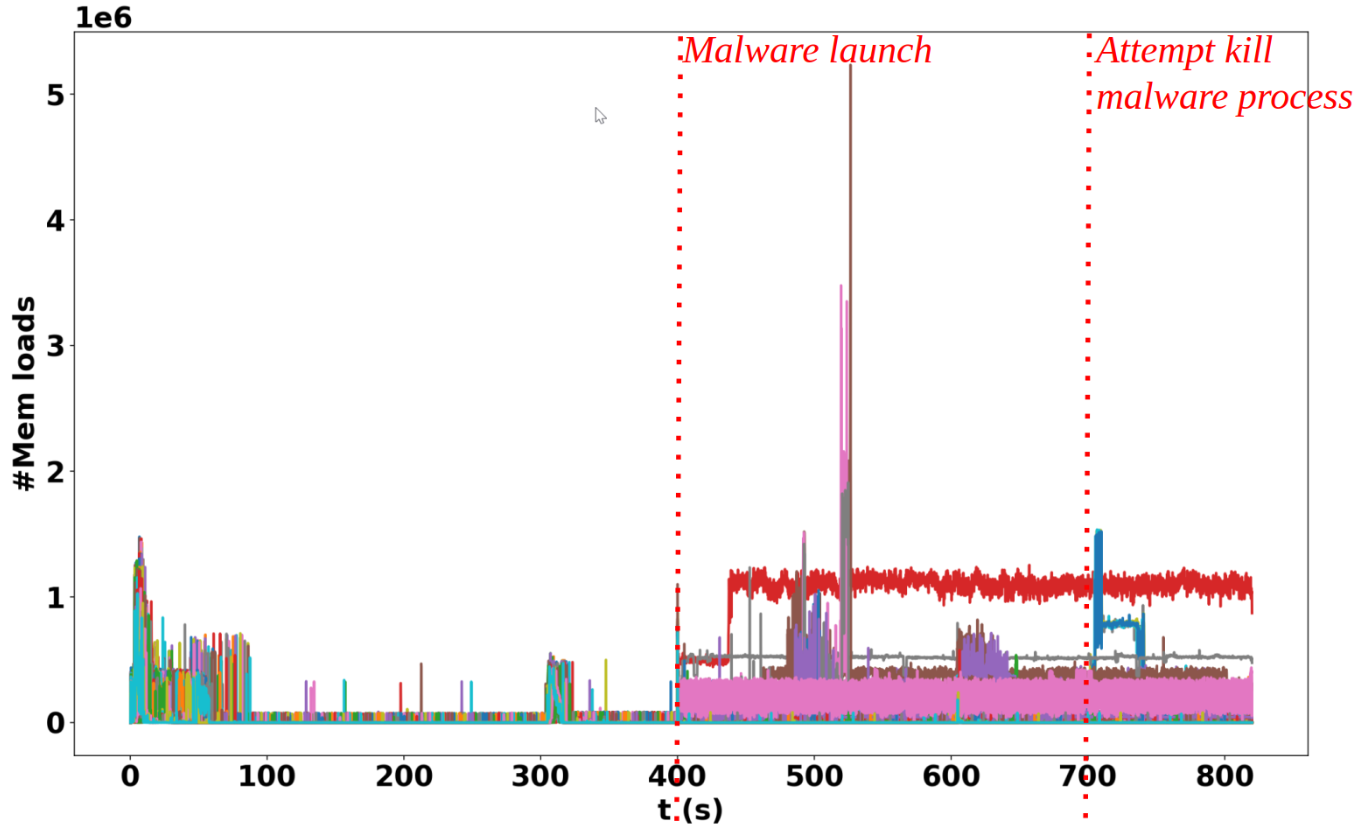}
}
\caption{Time series HPC measurements of numbers of instructions (top), branches
  (middle), and memory loads (bottom). Different colors correspond to measurements from different samples.}
\label{fig:measurements_HPC_instructions}
\end{figure}

\begin{figure}[!h]
\centerline{
  \includegraphics[width=0.9\linewidth]{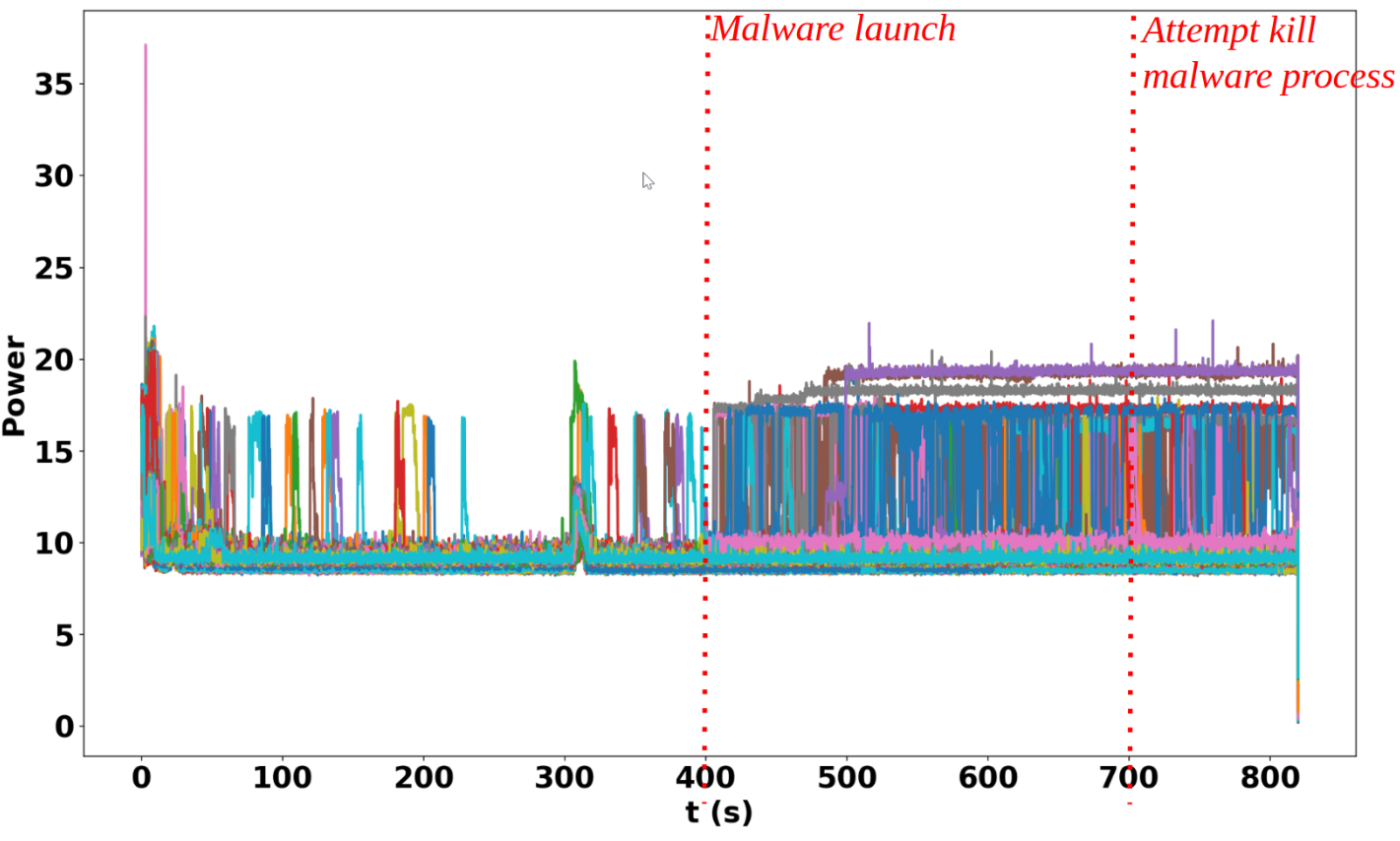}
}
\caption{Time series measurements of power usage. Different colors correspond to measurements from different samples.}
\label{fig:measurements_power}
\end{figure}

\begin{figure}[!h]
\centerline{
  \includegraphics[width=0.9\linewidth]{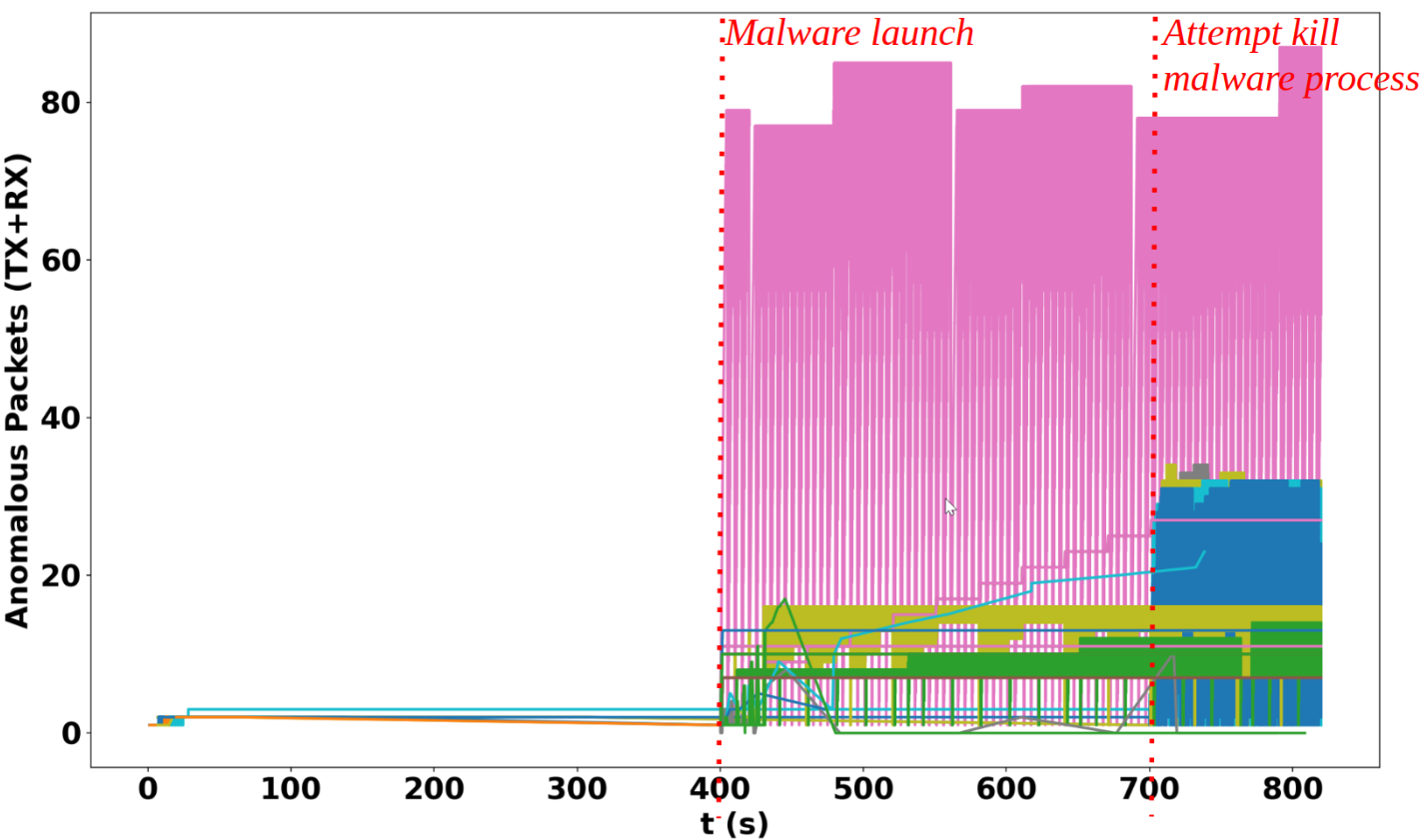}
}
\caption{Time series measurements of network activity (\# packets TX + RX); known IP addresses visible in no-malware baseline filtered for visual clarity. Different colors correspond to measurements from different samples.}
\label{fig:measurements_NIC}
\end{figure}

For example, sub component time series data collected for the elf.capoae malware sample from malpedia are shown in Figures~\ref{fig:capoae_nic} and \ref{fig:capoae_hpc}. 
To illustrate the variations of the time series measurements across different
malware samples\footnote{The malware samples in this set and the subset of 39 samples discussed later in the paper were provided by collaborators at Purdue (S. Jagannathan, X. Zhang). Other malware samples discussed in this paper are from malpedia.}, plots of several measurement modalities in the subcomponent
time series data are shown in
Figures~\ref{fig:measurements_HPC_instructions}-\ref{fig:measurements_NIC}. For example, in
Figure~\ref{fig:measurements_HPC_instructions}:Top, the time series of HPC
measurements of the numbers of instructions (over 100 ms time intervals) are
shown with different colors corresponding to measurements from different
malware samples. As discussed above, the malware sample is launched at $t=400$~s. It is
seen that several malware samples show observable apparent activity after the
malware launch at 400 s. Similarly, the plots for the HPC measurements of
numbers of branches and numbers of memory loads are shown in
Figures~\ref{fig:measurements_HPC_instructions}:Middle and
\ref{fig:measurements_HPC_instructions}:Bottom, respectively, and the plots for the
power measurements are shown in Figure~\ref{fig:measurements_power}. The plots
for the NIC measurements of numbers of packets transmitted/received (aggregated
over 50 ms time intervals) are shown in Figure~\ref{fig:measurements_NIC}, with
the known IP addresses visible in the no-malware baseline filtered for visual
clarity of the plots. Plots can be similarly drawn for the other subcomponent
measurements as well.

\begin{figure}[!h]
\centerline{
  \includegraphics[width=0.8\linewidth]{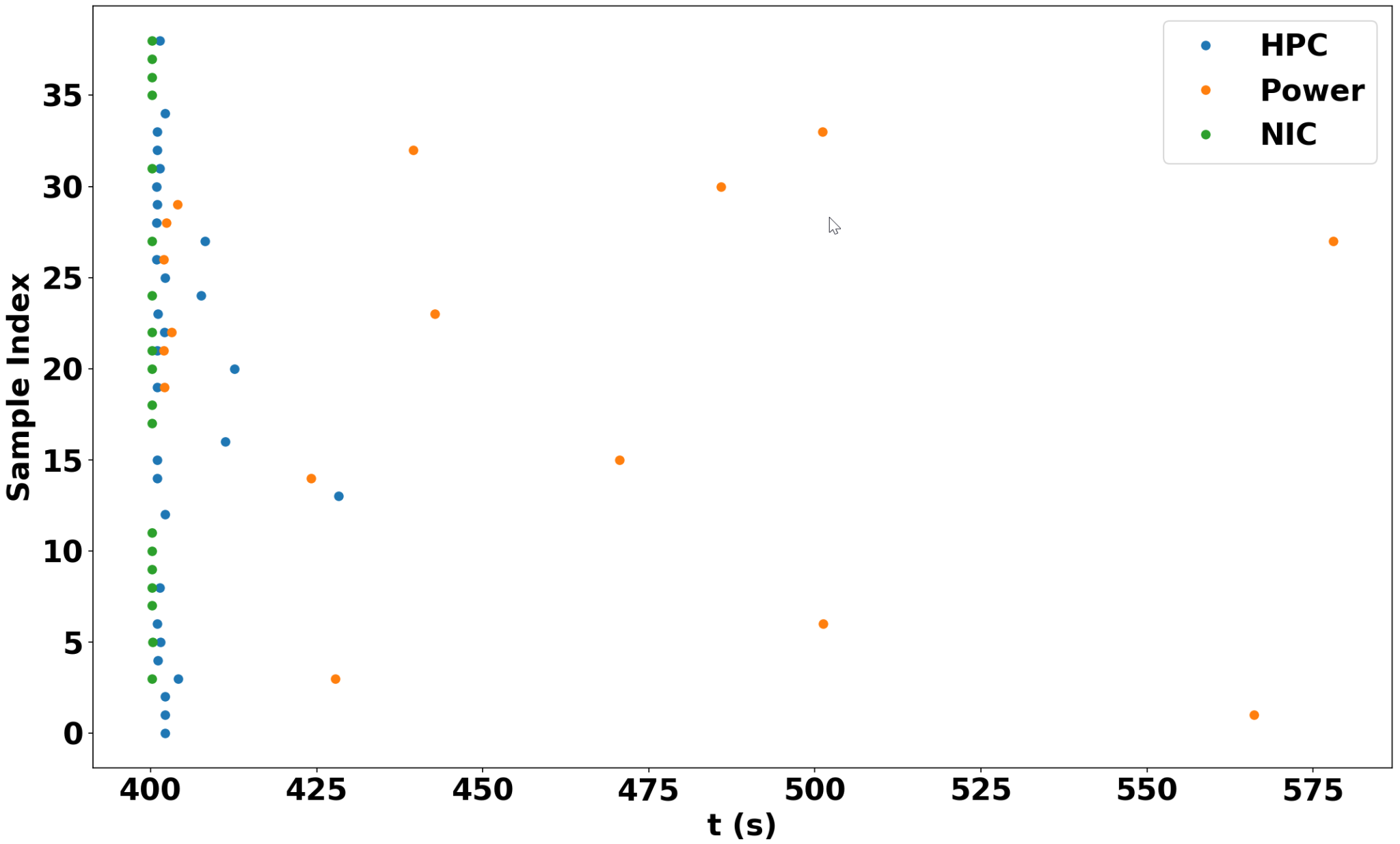}}
\centerline{  \includegraphics[width=0.8\linewidth]{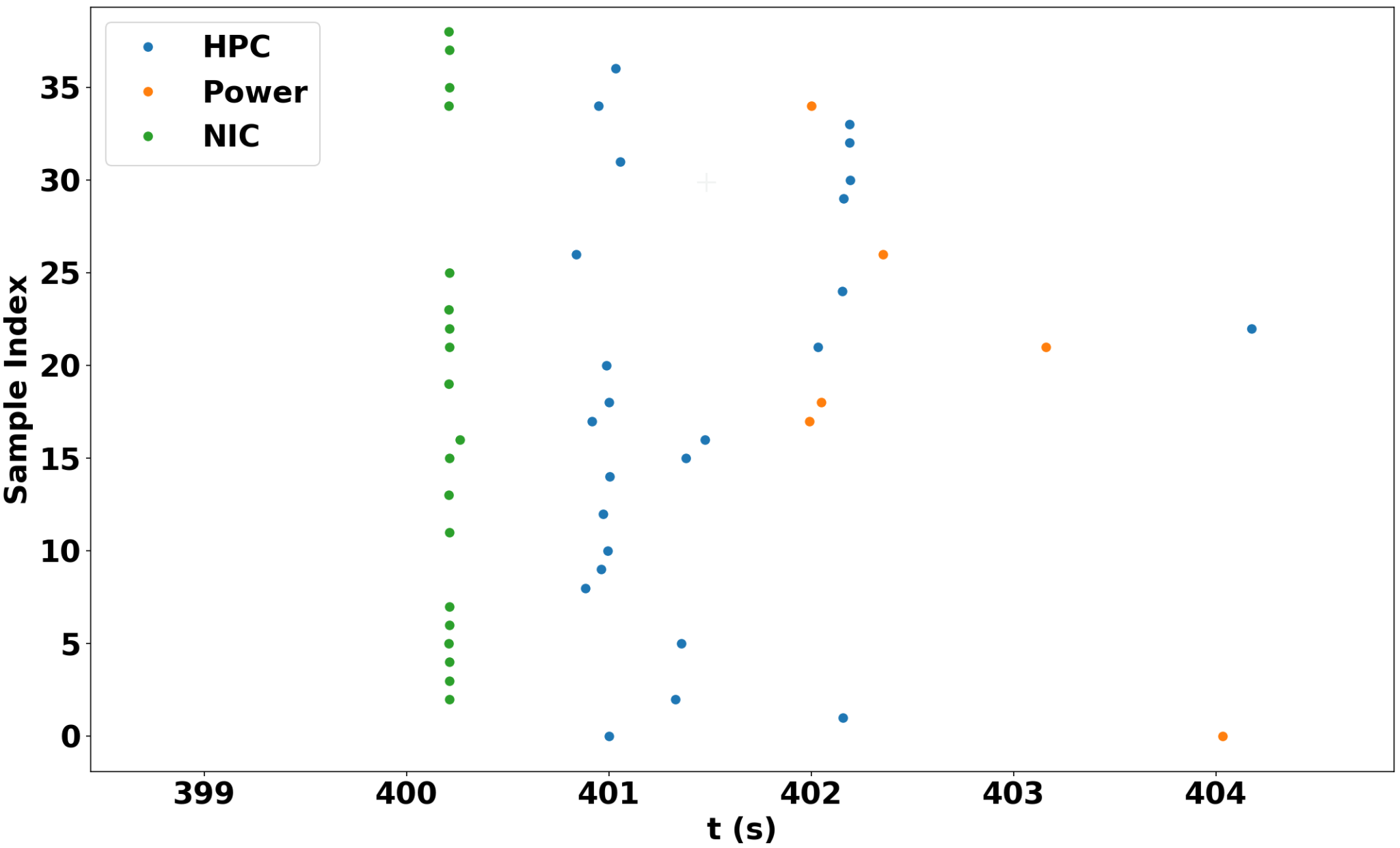}
}
\caption{Anomaly detections of subcomponent time series datasets collected when running different malware samples. The X-axis shows the time (in seconds) at which an anomaly was first detected. The malware samples are started at $t=400$~s. The Y-axis shows the sample index (each index is a different malware sample) for the 39 malware samples that were detected as having observable anomalous activity. The bottom plot shows a zoomed-in view. The blue, orange, and green dots correspond to anomaly detections using the HPC, power, and NIC measurements, respectively.}
\label{fig:anomaly_detection}
\end{figure}

Based on the subcomponent measurements outlined above, a machine learning (ML) based anomaly detector was implemented. The ML models are trained using only baseline data (i.e., without using any data collected with a malware sample). Hence, the ML models are one-class classifiers since they essentially model the ``good'' behavior and flag anything deviating the good behavior as anomalous.
An Isolation Forest model was used for the HPC measurements and a Local Outlier Factor model for the power measurements. A statistical traffic analysis model was used for the NIC measurements.  
For a set of 39 malware samples that show observable activity in the HPC, power, and NIC subcomponent measurements, Figure~\ref{fig:anomaly_detection} shows the times at which anomalous activity was first detected for each malware sample from the time series measurements from each of these subcomponents. The bottom plot in Figure~\ref{fig:anomaly_detection} shows a zoomed-in view.
The malware samples are started at $t=400$~s as discussed above. It is seen that for most of the malware samples, the most rapid detection is with the NIC or the HPC measurements, depending on the specific behaviors initiated by the malware. Since the power usage is essentially dependent on the system activity but can be viewed as a low-pass filtered version of the HPC measurements, the power measurements tend to be somewhat slower in detecting anomalies compared to the HPC and NIC measurements.    
Since different subcomponent measurements can expose different types of malware, the numbers of malware samples detected by the subcomponent measurements operating in isolation were also analyzed as summarized in Table~\ref{tab:malware_anomaly_detections}.
It is seen that the HPC measurements detected the most number of malware samples, followed by the NIC measurements, and then the power measurements. Since the power measurements are closely related to the same underlying physical activities for which the HPC measurements provide a more fine-grained view, the set of malware samples detected using the power measurements was seen to be a subset of the malware samples detected using the HPC measurements. On the other hand, depending on the malware behavior, some samples were detected only with the NIC measurements and not the HPC/power measurements or the other way around. The combination of the subcomponent measurements provides a more comprehensive view of the system operation and enables more robust anomaly detection. The GPU and keyboard time series measurements are not explicitly discussed above since these malware samples did not involve the GPU activity or keyboard interactions.

\begin{table}
  \caption{Numbers of malware samples detected using different combinations of subcomponent measurements.}
  \label{tab:malware_anomaly_detections}
  \centering
  \begin{small}
  \begin{tabular}{|c|c|}
    \hline
    Side Channel & \# samples detected \\
    \hline
    HPC & 30 \\
    Power & 16 \\
    NIC & 19 \\
    HPC+Power & 30 \\
    HPC+Power+NIC & 39 \\
    \hline
  \end{tabular}
  \end{small}
\end{table}

\begin{figure}[!h]
  \centerline{
    \includegraphics[width=\linewidth]{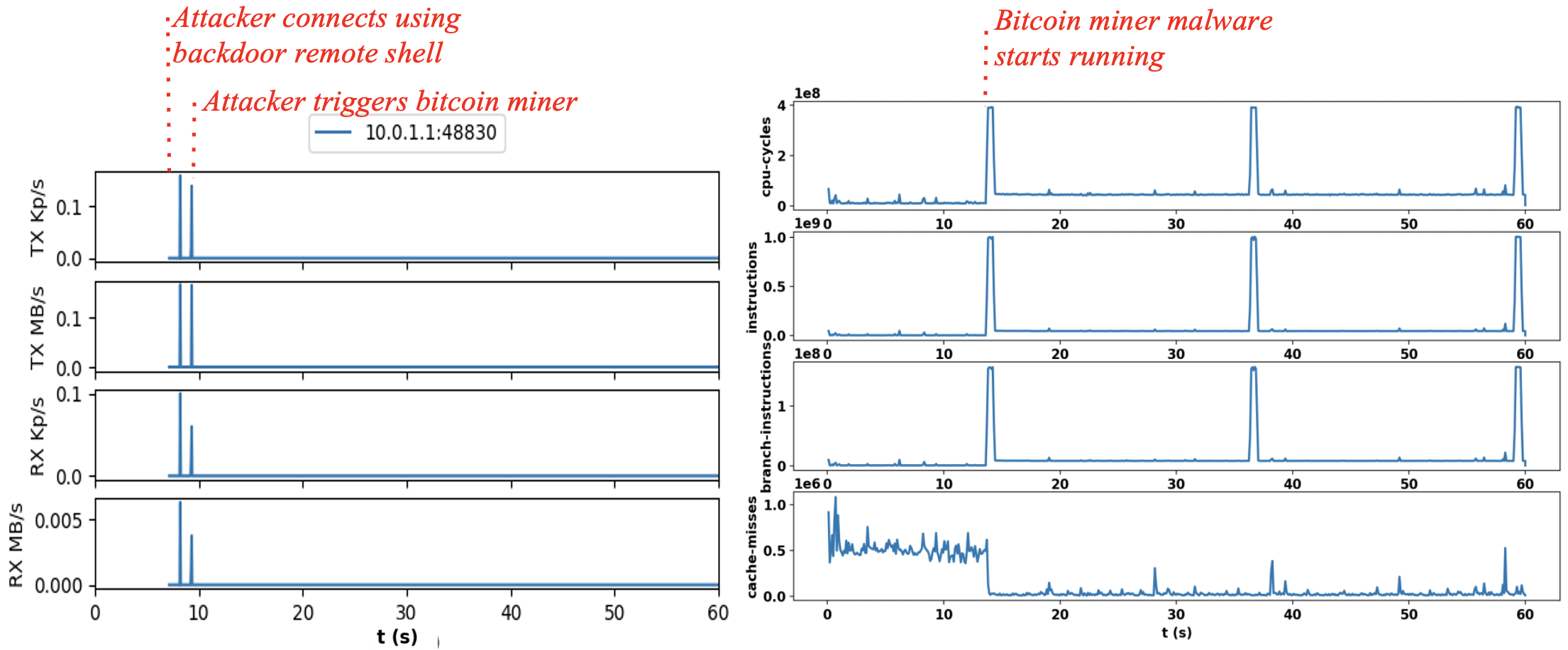}
  }
  \vspace*{-0.1in}
  \caption{NIC and CPU HPC time series measurements collected when malpedia/elf.rbs\_srv (a backdoor remote shell on local port 1234) is used by a remote attacker to connect to the system and trigger a bitcoin miner.}
  \label{fig:rbs_srv_miner}
  \vspace*{-0.1in}
\end{figure}

\begin{figure}[!h]
  \centerline{
    \includegraphics[width=\linewidth]{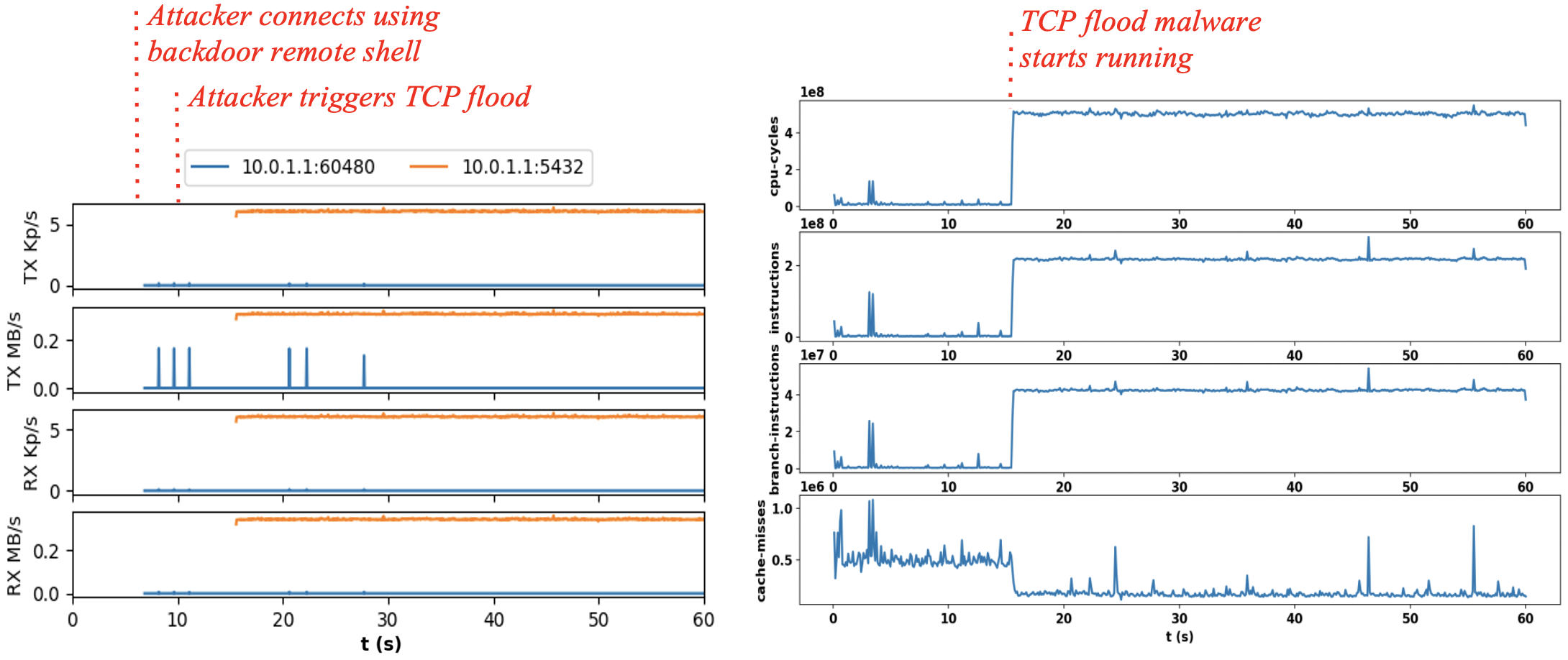}
  }
  \vspace*{-0.1in}
  \caption{NIC and CPU HPC time series measurements collected when
    malpedia/elf.rbs\_srv (a backdoor remote shell on local port 1234) is used by
    a remote attacker to connect to the system and trigger a TCP flood attack
    using malpedia/elf.floodor.}
  \label{fig:rbs_srv_flood}
  \vspace*{-0.1in}
\end{figure}

In addition to subcomponent time series measurement data collection for each malware sample as discussed above, sequences of adversary interactions involving multiple malware samples were also studied. In particular, sequences of adversary interactions including gaining unauthorized access through Remote Access Trojans (RATs) and launching processes such as cryptominers and flood attacks have been modeled to collect subcomponent measurement datasets under such interactions. RATs provide backdoor access to attackers via remotely accessible shells, which can then be used by the attacker to initiate successive actions such as deploying additional malware, running malicious commands, changing system configuration, attempting lateral movements, etc. To model representative sequences of attacker interactions, subcomponent measurements were collected for scenarios such as the attacker connecting using a RAT and triggering processes such as cryptominers or TCP/UDP flood attacks.
    For example, in Figure~\ref{fig:rbs_srv_miner}, the malpedia/elf.rbs\_srv RAT is used by a remote attacker to connect to the system and start running a cryptominer. The NIC measurements show the attacker's connection and activity. The Hardware Performance Counter (HPC) measurements on the CPU show the change in the system's activity when the cryptominer starts running. In Figure~\ref{fig:rbs_srv_flood}, the attacker connects using malpedia/elf.rbs\_srv and launches a TCP flood attack. The NIC measurements show both the initial connection using the rbs\_srv's backdoor and also the TCP flood (which is on a separate port). The CPU's HPC measurements show the system's activity including the process running the TCP flood.

\begin{figure}[!h]
  \centerline{
    \includegraphics[width=\linewidth]{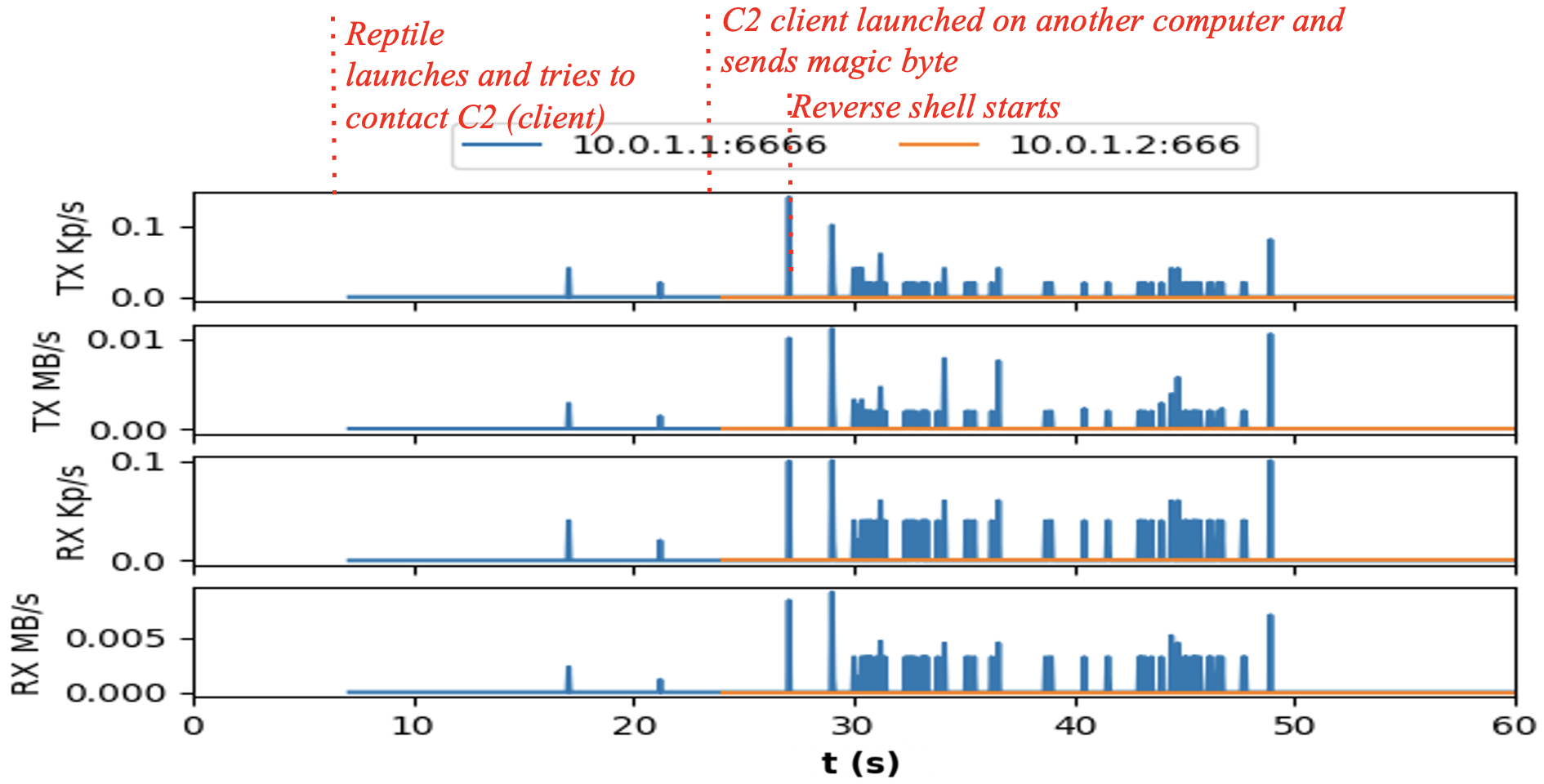}
  }
  \caption{NIC time series measurements shows sequence of interactions when
    malpedia/elf.reptile is launched and a remote attacker
    uses port knocking to get a reverse shell.}
  \label{fig:reptile}
\end{figure}

\begin{figure}[!h]
  \centerline{
    \includegraphics[width=\linewidth]{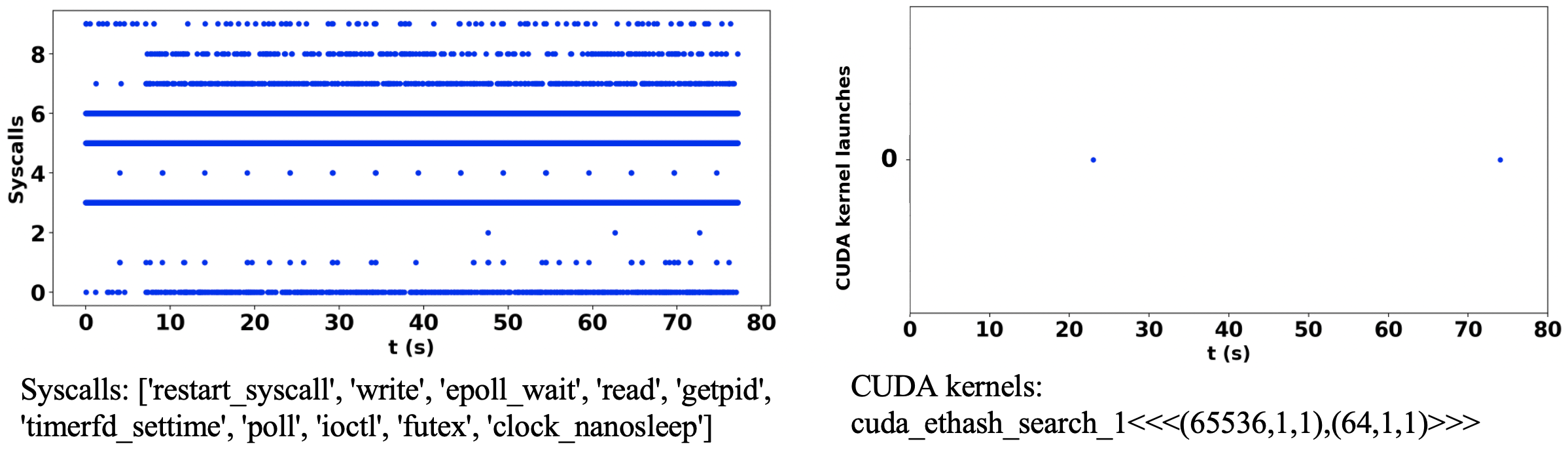}
  }
  \caption{System call and GPU CUDA kernel launch time series measurements when a
    GPU-based crypto-miner (T-Rex Ethereum) is triggered by a remote attacker via
    Reptile. The Y-axis corresponds to different detected system calls and CUDA
    kernel launches as shown below the corresponding plots.}
  \label{fig:eth_miner}
\end{figure}

    In Figure~\ref{fig:reptile}, the attacker uses the malpedia/elf.reptile RAT to connect to the system. When the RAT is launched, it initially tries to reach out to an external client (10.0.1.1 in this test). After that, it goes into a listening mode where it waits for a magic byte to arrive on a specific port. On receiving the magic byte, the RAT provides a reverse shell to the attacker who sent the magic byte. This sequence of activities is seen in the NIC measurements. The RAT uses a rootkit technique to hide its files, processes, kernel module, and network activity. While the network activity is not visible from host-level measurements since they are hidden by the rootkit, it is seen that firmware-level measurements from the NIC reveal the RAT's activity. On connecting to the system, the attacker can use the remote shell to launch other processes such as cryptominers and flood attacks resulting in subcomponent measurements similar to what was seen in the rbs\_srv tests in Figures~\ref{fig:rbs_srv_miner} and \ref{fig:rbs_srv_flood}. Also, subcomponent measurements for another cryptominer (T-Rex Ethereum) that runs cryptomining on the GPU are shown in Figure~\ref{fig:eth_miner}. The time series of system calls and CUDA kernel launches are shown in Figure~\ref{fig:eth_miner} when running this GPU cryptominer.

\appendices

\section{Filesystem-Aware Storage Activity Monitoring Using SATA and NBD}
\label{sec:sata_nbd}

\begin{figure}[!t]
\centerline{
\includegraphics[width=0.95\linewidth]{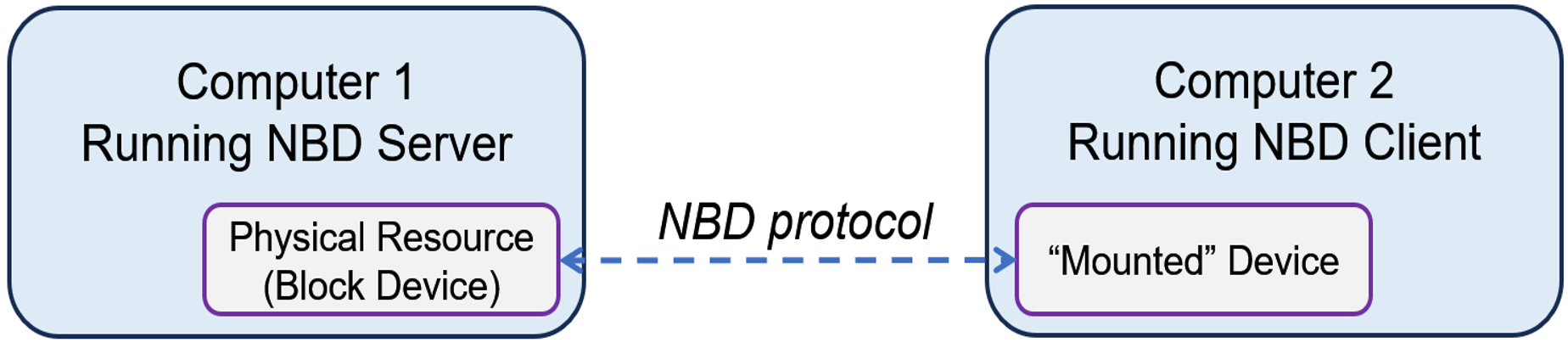}
}
\caption{Interfacing of block devices to the target machine via NBD.}
\label{fig:NBD_architecture}
\end{figure}

A prototype implementation of an FPGA-based SATA controller was developed to
explore the feasibility of real-time collection of SATA activity. Since SATA-based data collection
currently runs as part of a separate execution framework,
SATA measurements were not collected as part of Section~\ref{sec:subcomponents}. However, since this separate framework ({\bf SHIELD} -- Secure Host-Independent Extensible Logging for
SATA/Network Storage Towards Ransomware Detection) based on SATA
and NBD enables deep off-host tamper-proof filesystem-aware side channels that can provide complementary benefits
along with the other side channels in the dataset, this measurement framework is
briefly discussed in this appendix.
The prototype implementation used a Digilent XUPV5 (Virtex-5 XC5VLX110T) FPGA
and a SATA Host Bus Adapter (HBA) architecture based on the open-source
GroundHog SATA HBA by ETH Zurich and Microsoft Research.
The Digilent XUPV5 FPGA provides multiple useful features for communication and
SATA design: High Speed Gigabit Transceiver Ports (GTP) capable of achieving the
3 Gb/s throughput required for SATA generation 2 communication; on-board SATA hardware with dedicated integrated circuits for SATA clock generation for GTPs, as well as physical SATA connectors, enabling direct interfacing with storage devices;
extensive FPGA primitives including logic blocks, memory, and DSP primitives, allowing for expansion when implementing  detection circuitry.
The FPGA uses an Ethernet interface to communicate between a Host PC and the SATA HBA. The HBA communicates directly with a HDD/SSD at up to SATA Gen2 Speed (3 Gb/s)  using high-speed transceivers (GTPs) on the FPGA fabric.
While the initial GroundHog SATA HBA only supports a Windows host (with support
only for older Windows versions such as Windows 7)  using Microsoft's Simple
Interface for Reconfigurable Computing (SIRC),we developed a prototype
OS-agnostic control interface to remove requirements for Windows OS primitives
or software-side SIRC for communication.
Our OS-agnostic implementation uses a multithreaded mode to allow for simultaneous packet reception and transmission for evaluation of status information from received packets and generation of subsequent control logic commands.
Benchmark measurements of the prototype SATA implementation show high throughput when reading and writing sequentially, 
close to the SATA2 maximum throughput ($\sim$270 MiB/s), and zero SATA errors in the HBA when reading and writing at high volume.
Additionally, a mechanism was implemented to allow connecting the raw read/write
Ethernet-based SATA functions to an NBD interface so as to ``mount'' the HDD/SSD drive on the target machine and access the storage via filesystem drivers.
NBD is a block access protocol that enables exporting of a block device from one computer to another. 
The ``block device'' is a very generic concept and can be anything that allows reads and optionally writes. 
The block device can be an actual connected device, a file, or even just programmatically defined functions that dynamically implement conceptual ``reads'' and/or ``writes.''
The Ethernet-based raw SATA read/write interface can be mapped to the NBD API to allow mounting of the SATA drive on the target machine as shown in Figure~\ref{fig:NBD_architecture}. 
An NBD server on the machine to which the storage media (in this case, the FPGA-based SATA drive) is physically connected exports the storage media as a block device over the network. The NBD client running on the target machine connects to the NBD server and can access the storage media through read/write functions or can optionally mount the storage media as a block device.
Additionally, NBD monitoring can be potentially useful even apart from the SATA application. For example, the NBD API could be mapped to file images with different filesystems as well as to other block devices such as RAM disks. The plugin structure of the NBD API facilitates inserting of real-time monitoring of read/write operations irrespective of the storage back-end.

\begin{figure}[!pt]
            \centering
            \includegraphics[width=\linewidth]{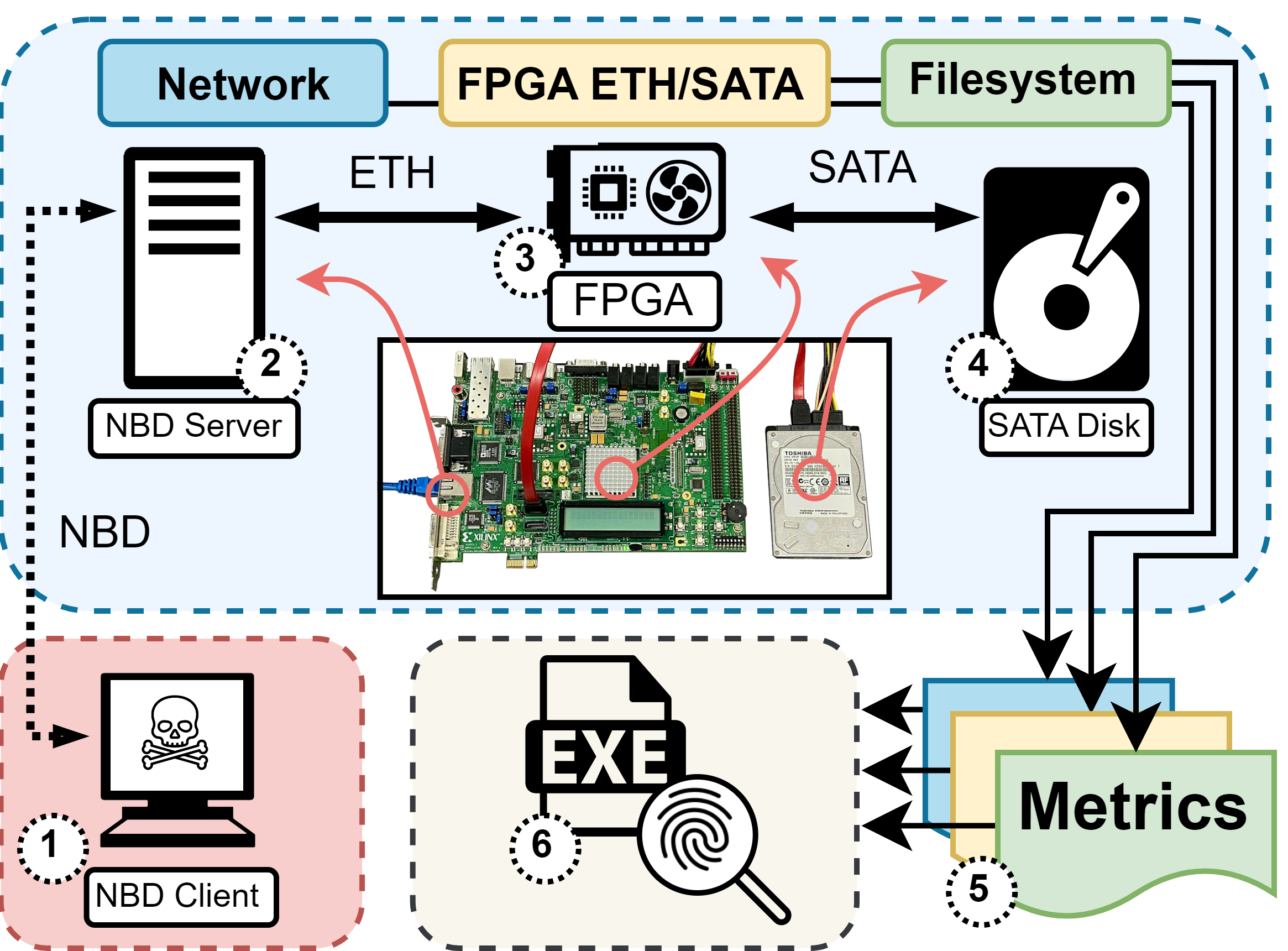}
            \caption{Architectural overview of \textsc{Shield} components and flow.}
            \label{fig:arch}
        \end{figure}

The primary objective of \textsc{Shield} is to create a cost-effective and
host-independent architecture for capturing metrics from the SATA, file system,
and server-side NBD layer, allowing the use of these metrics in automated,
real-time hardware based detection of malware-induced anomalous behavior. Fig.~\ref{fig:arch} shows an overview of this architecture.
The multi-layered approach of \textsc{SHIELD} provides not only fundamental
measurements such as access types and sizes, but also detailed metrics unique to
the file system. Although \textsc{Shield} can be modified to support different
file systems, our proof-of-concept considers the EXT4 filesystem. For EXT4, this architecture can capture unique metrics like access and modification events in the superblock, which stores key disk information such as usage and block sizes; group descriptor tables (GDT), which map disk features and usage patterns; inodes, which contain file metadata and pointers to data blocks; and inode data blocks, which hold the actual file data.
        The NBD interface architecture was designed to support both physical SATA devices and
        virtual disk images to facilitate experimentation and testing.

            The flow through \textsc{Shield} components is illustrated in
            Fig.~\ref{fig:arch}. It is initiated by an NBD client \circled{1} on
            the host, which may be a separate machine or a virtualized
            environment on the same system. This client connects to the NBD
            server \circled{2}, communicating over Ethernet to the FPGA
            \circled{3}, which handles measurement and logging of the desired
            metrics. The FPGA interfaces directly with a SATA disk \circled{4},
            enabling low-level access to data storage and file system
            structures. Metrics are collected across multiple layers—including
            network, FPGA, and file system levels—and then observed in real-time
            or recorded \circled{5} before changes are committed to the disk.
            These recorded metrics are inputs for behavior fingerprinting and
            anomaly detection, to distinguish between benign and malicious
            activities \circled{6}.  metrics are used to capture the file system interactions in real-time.
        
            The NBD server interfaces with the client-side host and uses either (a) an Ethernet packet generation library for FPGA communication or (b) access functions to interact with a local virtual-disk image. This setup allows the host—a separate machine or a virtualized instance—to perform standard file system operations (e.g., mounting, reading, writing) as though directly connected with a physical disk. The NBD server also captures metrics like read/write counts, access sizes, pending requests, and disk utilization to ensure consistency with hardware metrics. 

\begin{figure}[tp!]
        \centering
        \includegraphics[width=0.85\linewidth]{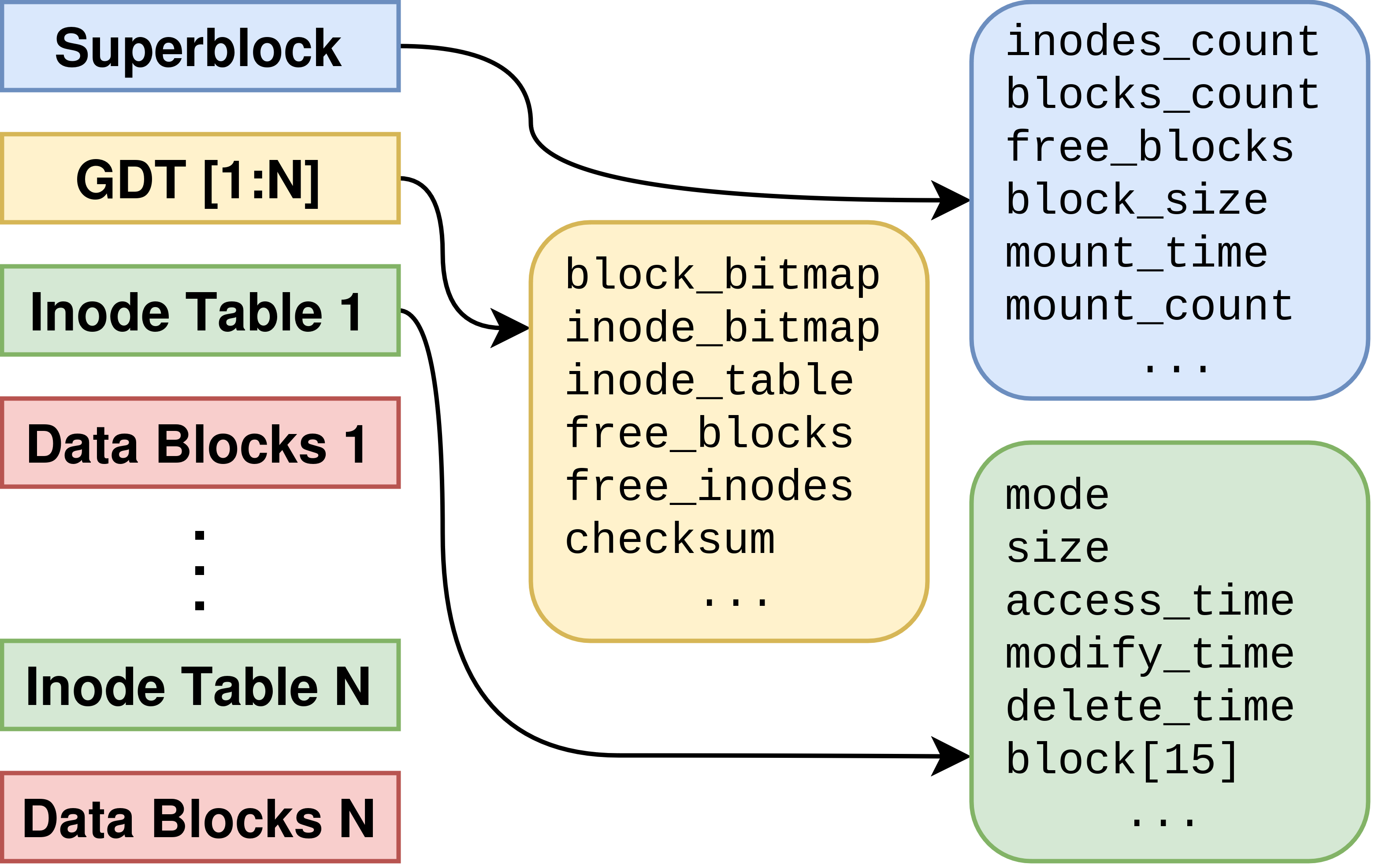}
        \caption{EXT4 architecture with metrics used in ransomware detection.}
        \label{fig:ext4} 
    \end{figure}

            The FPGA between the NBD server and storage device captures
            additional metrics for accurate software fingerprinting. Basic
            measurements include access type (read/write), access size, and disk
            location. With direct disk and data buffer access, the hardware can
            analyze detailed file system features, such as the superblock, group
            descriptor tables, and inodes, to track granular changes within the
            EXT4 filesystem. An overview of relevant EXT4 structures utilized by
            \textsc{Shield} is depicted in Fig.~\ref{fig:ext4}. Beyond these
            core metrics, the EXT4 filesystem offers additional data points,
            like journals and extent trees, though these are outside the current
            scope of \textsc{Shield}. The hardware can collect metrics in two
            ways: (a) actively, by using direct SATA commands to read data
            independently from the host, and (b) passively, by parsing buffers
            as data is read/written by the host. This dual approach allows the hardware to perform an initial file system scan during disk initialization, before the host enumerates the drive, and then switch to passive monitoring during regular host operations.

\bibliographystyle{IEEEtran}
\bibliography{refs}   

\end{document}